\documentclass[pre,amsmath,amssymb,reprint,aps]{revtex4-2}   
 
\usepackage{xcolor}
\definecolor{indigo}{RGB}{0,0,120}

\usepackage[skip=3pt]{subcaption}

\usepackage{ragged2e}
\DeclareCaptionJustification{justified}{\justifying}
\usepackage{graphicx}
\usepackage{soul}
\usepackage{cancel}
\usepackage{yfonts}
\usepackage{tikz}
\usetikzlibrary{calc}
\usetikzlibrary{decorations.markings}

\newcommand{\bra}{\langle}
\newcommand{\ket}{\rangle}

\newcommand{\tl}[1]{\tilde{#1}}

\newcommand{\pdr}{\partial}

\def\mod{\text{mod~}}

\newcommand{\beq}{\begin{equation}}
\newcommand{\eeq}{\end{equation}}
\newcommand{\beqs}{\begin{eqnarray}}
\newcommand{\eeqs}{\end{eqnarray}}

\newcommand{\half}{\frac{1}{2}}
\newcommand{\ov}[1]{\frac{1}{#1}}
\newcommand{\fr}[2]{\frac{#1}{#2}}

\newcounter{rowno}
\def\imply{\Rightarrow}

\def\fl{\noindent}

\def\al{\alpha}		
\def\g{\gamma} 		
\def\G{\Gamma} 
\def\del{\delta}

\def\sig{\sigma}

\def\Sig{\Sigma}		
\def\tht{\theta}	
		
\def\om{\omega}

\def\vphi{\varphi}
\def\vf{\varphi}

\usepackage{titlesec}
\titleformat{\section}{\normalsize\bfseries}{\thesection}{1em}{}
\titleformat{\subsection}{\small\bfseries}{\thesubsection}{1em}{}
\titleformat{\subsubsection}{\small\bfseries}{\thesubsubsection}{1em}{}

\newcommand{\Red}{\color{red}}
\newcommand{\Blue}{\color{blue}}

\newcommand{\Y}{\color{yellow}}

\newcount\colveccount  
\newcommand*\colvec[1]{\global\colveccount#1  \begin{pmatrix} \colvecnext} \def\colvecnext#1{#1 \global\advance\colveccount-1
        \ifnum\colveccount>0 \\ \expandafter\colvecnext
        \else \end{pmatrix} \fi}

\newenvironment{smmat}
  {\left(\begin{smallmatrix}}
  {\end{smallmatrix}\right)}

\usepackage{calligra} 
\DeclareMathAlphabet{\mathcalligra}{T1}{calligra}{m}{n}
\DeclareFontShape{T1}{calligra}{m}{n}{<->s*[2.2]callig15}{}

\newcommand{\longsquiggly}{\xymatrix{{}\ar@{~>}[r]&{}}}

\newcommand{\tlg}{\tilde g}

\newcommand{\mZ}{{\mathbb{Z}}}

\usepackage[colorlinks=true, linkcolor=indigo, citecolor=blue, urlcolor=indigo]{hyperref}

\begin{document}



\title[Quantum three rotor problem]{\hfill \href{https://arxiv.org/abs/2407.15482}{\tt{arXiv:2407.15482}} \\
Quantum three-rotor problem in the identity representation}

\author{Govind S. Krishnaswami$^{a}$}
\email{govind@cmi.ac.in}
\author{Himalaya Senapati$^{a, b}$}
\email{himalay@cmi.ac.in}
\affiliation{\small $^{a}$Physics Department, Chennai Mathematical Institute, SIPCOT IT Park, Siruseri 603103, India \\ 
\small $^{b}$HSBC, RMZ Ecospace Business Park, Bellandur, Bengaluru 560103, India}

\date{December 7, 2024}


\begin{abstract}

The quantum three-rotor problem concerns the dynamics of three equally massive particles moving on a circle subject to pairwise attractive cosine potentials and can model coupled Josephson junctions. Classically, it displays order-chaos-order behavior with increasing energy. The quantum system admits a dimensionless coupling with semiclassical behavior at strong coupling. We study stationary states with periodic `relative' wave functions. Perturbative and harmonic approximations capture the spectrum at weak coupling and that of low-lying states at strong coupling. More generally, the cumulative distribution of energy levels obtained by numerical diagonalization is well-described by a Weyl-like semiclassical estimate. However, the system has an $S_3 \times \mathbb{Z}_2$ symmetry that is obscured when working with relative angles. By exploiting a basis for invariant states, we obtain the spectrum restricted to the identity representation. To uncover universal quantum hallmarks of chaos, we partition the spectrum into energy windows where the classical motion is regular, mixed or chaotic and unfold each separately. At strong coupling, we find striking signatures of transitions between regularity and chaos: spacing distributions morph from Poisson to Wigner-Dyson while the number variance shifts from linear to logarithmic behavior at small lengths. Some nonuniversal features are also examined. For instance, for strong coupling, the number variance saturates and oscillates at large lengths while the spectral form factor displays a nonuniversal peak at short times. Moreover, deviations from Poisson spacings at asymptotically low and high energies are well-explained by quantum harmonic and free-rotor spectra projected to the identity representation at strong and weak coupling. Interestingly, the degeneracy of free-rotor levels admits an elegant formula that we deduce using properties of Eisenstein primes.

\end{abstract}

\maketitle


\small

\tableofcontents

\normalsize

\section{Introduction}
\label{s:introduction}

The study of quantum manifestations of chaos and the transition from regular to chaotic behavior in Hamiltonian systems has been a fertile area of research with several applications such as to the electronic states of a donor impurity in a semiconductor crystal \cite{gutzwiller-book} and to quantum dynamical localization and transport \cite{santhanam-phys-rpts}. The development of quantum chaology has benefited from the study of several model systems such as billiard problems \cite{stockmann}, the anisotropic Kepler problem \cite{gutzwiller-book}, free particle motion on constant negative curvature Riemann surfaces \cite{balazs-voros-chaos-pseudosphere}, kicked rotors and tops \cite{haake} and the planar elastic pendulum \cite{lowenstein}. Some themes that have emerged from these studies in quantifying quantum chaos are energy level spacing statistics (level repulsion and avoided crossings) \cite{bohigas-giannoni-schmit}, trace formulae and sums over periodic orbits as a bridge between classical and quantum mechanics \cite{gutzwiller-book}, the role of discrete symmetries in relating spectral statistics to predictions from random matrix ensembles \cite{keating-gue-goe, lss-gue-billiard-1996, joyner-muller-sieber-2012}, the nature of Wigner and Husimi functions \cite{tabor-chaos-integ-book}, nodal patterns of wave functions \cite{jain-samajdar-rmp} and out-of-time-ordered correlators \cite{larkin, shenker-stanford}. These investigations have revealed both universal features (modeled by random matrix ensembles) depending only on the symmetry class of the model as well as system-dependent features (such as oscillations in the spectral rigidity) on various scales \cite{berry-spec-rigidity}. 

In this paper, we focus on the quantum three-rotor system that models a chain of coupled Josephson Junctions \cite{sondhi-girvin, shnirman-small-JJ, orlando-mooij-persis-curr-qubit, govind-ankit-bifurcation}. Interestingly, the classical version has also been found to display rich features \cite{gskhs-3rotor,gskhs-3rotor-ergodicity}. In the classical three-rotor problem, point particles of mass $m$ move on a circle of radius $r$ subject to attractive cosine interparticle potentials. If the rotor angles are denoted $\tht_{1,2,3}$, then their total energy is
	\beqs
	E_{\rm cl} &=& \frac{m r^2}{2}(\dot \tht_1^2 + \dot \tht_2^2 + \dot \tht_3^2)
	+ g [3 - \cos(\tht_1-\tht_2) \cr
	&& - \cos(\tht_2-\tht_3) - \cos(\tht_3-\tht_1) ]. \; \quad
	\eeqs
In terms of the center of mass angle $\vf_0 = (\tht_1+\tht_2+\tht_3)/3$ and relative angles $\vf_{1} = \tht_1 - \tht_2$ and $\vf_2 = \tht_2 - \tht_3$, the energy of relative motion is 
	\beqs
	E &=& \fr{mr^2}{3} \left( \dot \vf_1^2 + \dot \vf_2^2 + \dot \vf_1 \dot \vf_2 \right) \cr
	&+& g [3 - \cos \vf_1 - \cos \vf_2 - \cos(\vf_1 + \vf_2) ].
	\eeqs
In the application to coupled Josephson junctions, rotor angles correspond to superconducting phases of metallic segments \cite{govind-ankit-bifurcation}. Quite apart from this physical realization, the three-rotor problem is interesting for its rich dynamics. For instance, it displays order-chaos-order behavior with increasing relative energy $E$ and a globally chaotic phase ($5.33g \lesssim E \lesssim 5.6g$) where the dynamics has been argued to be ergodic and mixing \cite{gskhs-3rotor-ergodicity}. Intriguingly, the transition to widespread chaos around $E = 4 g$ seems to be associated with (i) an accumulation of a doubly infinite sequence of fork-like isochronous and period-doubling bifurcations \cite{govind-ankit-bifurcation} in the pendulum family of periodic orbits and (ii) a loss of strict positivity in the curvature of the Jacobi-Maupertuis metric on the configuration space \cite{gskhs-3rotor}. Several other remarkable phenomena concerning the classical three-rotor problem have been reported in Refs.~\cite{gskhs-3rotor,gskhs-3rotor-ergodicity,govind-ankit-bifurcation}. 

A natural question concerns the quantum dynamics of three rotors and in particular, quantum manifestations of its order-chaos-order behavior. In this paper, we initiate the study of the quantum three-rotor problem and address this question by examining the spectrum of stationary states and investigate both universal as well as system-specific properties. We begin in Section \ref{s:formulation} by formulating the quantum theory via canonical quantization. The parameters $m, r$ and $g$ of the classical system do not admit any dimensionless combination. Thus, the energy $E$ of a state in units of $g$ served as a dimensionless classical control parameter, with order-chaos-order behavior with increasing energy \cite{gskhs-3rotor}. A distinctive feature of the quantum system is the possibility of using Planck's constant to define a dimensionless coupling $\tlg = g m r^2/\hbar^2$, which helps in organizing the quantum theory. The weak coupling limit $\tlg \to 0$ is an extreme quantum limit while semiclassical behavior may be expected when $\tlg$ is large. In Section \ref{s:SOV-BC}, we show that the passage from rotor angles to center of mass and relative angles also entails new features following from the requirement that the wave function be singlevalued. The center of mass and relative wave functions arising from separation of variables in the Schr\"odinger eigenvalue problem must satisfy any one of three `periodicity up to a phase' boundary conditions, labelled by the cube roots of unity. In the sequel, we restrict to the simplest type of boundary condition and the corresponding sector with center of mass angular momentum quantum number $\ell$ divisible by $3$. Although the use of relative angles reduces the number of degrees of freedom, it obscures an $S_3 \times \mathbb{Z}_2$ permutation-parity symmetry of the model. In Section \ref{s:sym-of-3rotor}, we identify the action of this group on relative angles. It will be seen to play a crucial role in decomposing the energy spectrum into irreducible representations.

Section \ref{s:approx-results} contains results that employ variational, perturbative and harmonic approximations. Our variational ansatz accurately captures the energy of the ground state over the entire range of values of $\tlg$ from weak to strong coupling. Perturbation theory reveals the pattern of energy level degeneracies and their breaking at weak coupling. By exploiting number theoretic properties of the ring of Eisenstein integers, we  derive a closed form expression for the unperturbed level degeneracies in terms of prime factorization and explain why (with the exception of the ground state) they are multiples of $6$  (details are provided in Appendix \ref{a:degeneracy-formula-unperturbed}). On the other hand, our harmonic approximation around the ground state retains the $S_3 \times \mathbb{Z}_2$ symmetry and works well for a significant number of low lying states at strong coupling.

In Section \ref{s:numerical-diagonalization}, we begin by describing a spectral method for numerical diagonalization of the relative Hamiltonian in a Fourier basis. We find that, aside from the very low-lying states, the cumulative distribution of energy levels is closely captured by a Weyl-like semiclassical approximation that we derive. To examine spectral statistics and its universality, we decompose the Hilbert space into unitary irreducible representations of the $S_3 \times \mathbb{Z}_2$ symmetry group resulting in a partitioning of the spectrum. This process is often referred to as `desymmetrization' or occasionally `purification' though the latter term may be preferable since each invariant subspace continues to carry an action of the symmetry group. After enumerating the above representations in Section \ref{s:purify-spectrum}, we restrict attention to the identity representation. A basis for the invariant states carrying this representation is constructed and the matrix elements of the Hamiltonian are found in this basis. Numerical diagonalization of a truncated Hamiltonian of size $\sim 3 \times 10^4$ in this basis allows us to reliably determine the lowest $\sim 10^4$ levels and their spacings.

Having obtained the spectrum in the identity representation, we examine the statistics of nearest neighbor spacings in Section \ref{s:spacing-distrib}. To look for quantum manifestations of the classical order-chaos-order behavior with increasing energy, we first consider the semiclassical regime of large $\tlg \gtrsim 1000$. Here, we partition the spectrum into energy windows where the classical dynamics is regular, mixed and chaotic, and unfold each separately to have approximately unit mean spacing. This enables us to glimpse striking universal signatures of the transitions between regularity and chaos that were hidden in the unpartitioned spectrum: spacing distributions morph from Poisson to Wigner-Dyson (Gaussian orthogonal ensemble, since the Hamiltonian is time-reversal invariant and real symmetric) and back. Although the classical dynamics is regular at low energies, the low lying spectrum in the semiclassical regime should not be expected to show Poisson statistics. This is because the low energy dynamics may be approximated by a 2d harmonic oscillator. In fact, the spectral statistics in this regime is accurately captured by a desymmetrized or purified version of our harmonic approximation (Section \ref{s:harmonic-approx-purification-spacing}). On the other hand, in the weak coupling limit of small $\tlg$, we see departures from universal semiclassical expectations. As a matter of fact, in Section \ref{s:low-g-free-rotor}, we see signatures of the purified free-rotor spectrum superposed on the semiclassical Poisson distributions of spacings at high energies.

Section \ref{s:number-var-SFF} is devoted to two other spectral statistics: the number variance $\Sig(L)$ and the spectral form factor $K(\tau)$, both of which are two-point correlations in the energy spectrum. The number variance is the variance in the number of unfolded levels in a spectral window of length $L$. The variance is computed via an ensemble average over the centers of such spectral windows. As with level spacings, we find that evaluating the number variance in energy windows where the classical dynamics is chaotic or regular facilitates interpretation and comparison with universal expectations for small $L$ from the Gaussian orthogonal ensemble and a Poisson process. We find that for relatively large $\tlg$, the number variance in classically regular (for $\tlg = 300$) and chaotic (for $\tlg = 6000$) energy windows shows the expected linear and logarithmic behavior for small $L$ and then displays saturation and nonuniversal oscillations when $L$ is larger. In a similar vein, we see both universal and nonuniversal features in the spectral form factor, which also shows marked differences between regular and chaotic energy windows. We conclude in Section \ref{s:discussion} with a brief discussion of our results and open questions and conjectures inspired by our work.

\section{Formulation}
\label{s:formulation}

The classical Hamiltonian for the equal-mass three rotor system is
	\beq
	H_{\rm cl} = \sum_{i =1}^3  \left\{ \fr{\pi_i^2}{2 m r^2} + g [1- \cos\left(\tht_i-\tht_{i+1} \right) ] \right\}
	\label{e:classical-full-Hamiltonian}
	\eeq
with $\tht_1 \equiv \tht_4$ and $g > 0$ the strength of the attractive coupling. Here, $\tht_i$ are the rotor angles and $\pi_i$ their conjugate angular momenta: $\{\tht_i, \pi_j \} = \delta_{ij}$ \cite{gskhs-3rotor}. Quantizing canonically  by taking $\pi_i = -i \hbar \pdr_{\tht_i}$, we get the quantum Hamiltonian
	\beq
	H_{\rm tot} = \sum_{i=1}^3 - \fr{\hbar^2}{2 m r^2} \fr{\pdr^2}{\pdr\tht_i^2} + g [1- \cos(\tht_i - \tht_{i+1})].
	\eeq

\subsection{Center of mass and relative variables}

As in the classical theory\cite{gskhs-3rotor}, it is convenient to introduce center of mass (CM) and relative angles
	\beq
	\colvec{3}{\vf_0}{\vf_1}{\vf_2} = \colvec{3} { 1/3 & 1/3 & 1/3 } {1 & -1 & 0} {0 & 1 & -1} \colvec{3}{\tht_1}{\tht_2}{\tht_3},
	\label{e:cm-rel-ang-from-rotor-ang}
	\eeq
in terms of which $H_\text{tot} = H_{\rm CM} + H$ where
	\beqs
	H_{\rm CM} &=& - \fr{\hbar^2}{6 m r^2} \pdr^2_{\vf_0} \quad \text{and} \quad H = \fr{\hbar^2}{m r^2} T + g V(\vf_1, \vf_2) \cr 
	\text{with} \;\;
	T &=& - \left( \pdr^2_{\vf_1} - \pdr_{\vf_1} \pdr_{\vf_2} + \pdr^2_{\vf_2} \right) \cr
	\text{and}  \;\;  
	V &=& - \left[ \cos \vf_1 + \cos \vf_2 + \cos (\vf_1 + \vf_2) - 3 \right]. \quad
	 \label{e:Hamiltonian-total-CM-rel}
	\eeqs
Here, the conjugate momenta are represented by $p_j = -i \hbar \pdr_{\vf_j}$ for $j = 0, 1$ and $2$. 

\subsection{Nondimensional coupling and Hamiltonian}

The classical three-rotor model has three parameters $m,g$ and $r$. Since their dimensions are independent, the classical theory does not admit any dimensionless parameter. However, physical quantities may be nondimensionalized by measuring them in units of suitable combinations of $m,r$ and $g$. For instance, energies of states may be measured in units of $g$ and time in units of $\sqrt{mr^2/g}$. On the other hand, the quantum theory admits one dimensionless parameter due to the introduction of Planck's constant. We may take this dimensionless parameter to be the coupling constant $\tl g = mr^2 g/ \hbar^2$. Thus, upon nondimensionalization, measured values of physical quantities (eigenvalues of observables) become dimensionless functions of $\tl g$. Two limiting values of $\tl g$ are noteworthy. The weak coupling limit or extreme quantum limit is the one where $\tl g \to 0$. The strong coupling limit $\tl g \to \infty$ is a semiclassical limit.

Bearing this in mind, it is convenient to define a dimensionless Hamiltonian. In fact, there are two obvious possibilities
	\beq
	\tl H = \fr{H}{g} = \ov{\tl g} T + V \quad \text{and} \quad
	\hat H = \fr{H}{\hbar^2/mr^2} = T + \tl g V,\quad
	\label{dimless-H}
	\eeq
which are related by $\hat H = \tlg \tl H = (\tl g/g) H$. We will denote the eigenvalues of $H$, $\hat H$ and $\tl H$ by $E$, $\hat E$ and $\tl E$. Similarly, $\hat E_{\rm CM}$ denotes eigenvalues of $\hat H_{\rm CM}$.

\subsection{Separation of variables and boundary conditions}
\label{s:SOV-BC}

The boundary conditions (BC) on the wave function $\Psi$ in the CM and relative variables follow from the requirement that it be single-valued when viewed as a function of the $2\pi$-periodic rotor angles $\tht_{1,2,3}$. For instance, from (\ref{e:cm-rel-ang-from-rotor-ang}), $\tht_1 \mapsto \tht_1 + 2 \pi$ implies that $\vf_0 \mapsto \vf_0 + 2\pi/3$ and $\vf_1 \mapsto \vf_1 + 2\pi$. Thus, we have the following `boundary' conditions on $\Psi(\vf_{0,1,2})$:
	\beqs
	\Psi(\vf_{0,1,2})
	&=& \Psi(\vf_0 \pm 2\pi/3, \vf_1 \pm 2\pi, \vf_2) \cr
	&=& \Psi(\vf_0 \pm 2\pi/3, \vf_1, \vf_2 \mp 2\pi) \cr
	&=& \Psi(\vf_0 \pm 2\pi, \vf_1, \vf_2).
	\label{e:bcs-on-3torus}
	\eeqs
Since $H_{\rm tot}$ is a sum of CM and relative energies, we may factorize $\Psi(\vf_0, \vf_1, \vf_2) = \Upsilon(\vf_0)  \psi(\vf_1, \vf_2)$ in the Schr\"odinger eigenvalue problem $H_{\rm tot} \Psi = E_{\rm tot} \Psi$. We now derive BCs on $\psi(\vf_1, \vf_2)$ and $\Upsilon(\vf_0)$ by imposing (\ref{e:bcs-on-3torus}). To begin with, (\ref{e:bcs-on-3torus}) implies
	\beqs
	\Upsilon(\vf_0) &=& \Upsilon(\vf_0 \pm 2\pi) \quad \text{and} \cr
	\Upsilon(\vf_0) \psi(\vf_1, \vf_2) &=& \Upsilon(\vf_0 \pm 2\pi/3) \psi(\vf_1 \pm 2\pi, \vf_2) \cr
	&=& \Upsilon(\vf_0 \pm 2\pi/3) \psi(\vf_1, \vf_2 \mp 2\pi).
	\label{e:bcs-seprt-variable-on-3torus}
	\eeqs
Now, without loss of generality, we may write	
	\beq
	\psi(\vf_1 + 2\pi, \vf_2) = \varpi_1(\vf_1, \vf_2) \psi(\vf_1, \vf_2),
	\eeq
for some function $\varpi_1(\vf_1, \vf_2)$. This by (\ref{e:bcs-seprt-variable-on-3torus}) implies
	\beq
	\Upsilon(\vf_0 + 2\pi/3) = \varpi_1(\vf_1, \vf_2)^{-1}\Upsilon(\vf_0).
	\label{e:upsilon-varpi1}
	\eeq
Thus $\varpi_1$ must be a constant which again by (\ref{e:bcs-seprt-variable-on-3torus}) must be a cube root of unity:
	\beq
	\Upsilon(\vf_0) = \Upsilon(\vf_0+2\pi) \quad \imply \quad \varpi_1^3 = 1.
	\eeq 
Similarly, we may write  
	\beq
	\psi(\vf_1, \vf_2 - 2\pi) = \varpi_2(\vf_1, \vf_2) \psi(\vf_1, \vf_2)
	 \eeq
which by (\ref{e:bcs-seprt-variable-on-3torus}) implies
	\beq
	 \Upsilon(\vf_0 + 2\pi/3) = \varpi_2^{-1}\Upsilon(\vf_0).
	 \label{e:upsilon-varpi2}
	\eeq
From (\ref{e:upsilon-varpi1}) and (\ref{e:upsilon-varpi2}), $\varpi_1 = \varpi_2 \equiv \varpi$. In summary, the BCs on the CM and relative wave functions are
	\beqs
	\Upsilon(\vf_0 + 2\pi/3) &=& \varpi^{-1} \Upsilon(\vf_0) \quad \text{and} \cr
	\psi(\vf_1 + 2\pi, \vf_2) &=& \psi(\vf_1, \vf_2 - 2\pi) = \varpi \psi(\vf_1, \vf_2)
	\label{e:BCs-3torus}
	\eeqs
where $\varpi$ must be a cube root of unity. Hence, we arrive at the CM and relative Schr\"odinger eigenvalue problems which must be solved subject to (\ref{e:BCs-3torus}):
	\beq
	- \ov{6} \pdr^2_{\vf_0} \Upsilon = \hat E_{\rm CM}\Upsilon 
	\quad \text{and} \quad
	\left[ T + \tl g V \right] \psi = \hat E \psi
	\eeq
with $\hat E_{\rm tot} = \hat E_{\rm CM} + \hat E$. Although the BCs for $\varpi \neq 1$ render $\Upsilon$ and $\psi$ discontinuous (say at 0, $\pm 2\pi/3$ on the $\vf_0$-circle and along the two cycles $\vf_{1,2} = 0$ on the $\vf_1$-$\vf_2$ torus), the total wave function $\Psi$ is nevertheless continuous.

The center of mass eigenvalue problem is particularly simple with the solutions
	\beq
	\Upsilon(\vf_0) = e^{i \ell \vf_0} \quad \text{and} \quad \hat E_{\rm CM} = \ell^2/6
	\eeq
where $\ell$ is an integer with $\ell \equiv 0,1,2$ $(\mod 3)$ according as $\varpi = 1$, $e^{4\pi i/3}$ and $e^{2 \pi i/3}$. All CM energy levels other than $E_{\rm CM} = 0$ are doubly degenerate (clockwise and counterclockwise rotation of the CM corresponding to $\ell$ and $-\ell$) and is a consequence of the parity symmetry $\tht_i \mapsto - \tht_i$. For simplicity, we will henceforth restrict attention to the case $\varpi = 1$ and take $\ell = 0$ so that the CM is at rest. In this case, the BC on $\psi(\vf_1, \vf_2)$ reduces to $2 \pi$ periodicity in each relative angle.

\subsection{Symmetries of the three-rotor system}
\label{s:sym-of-3rotor}

Simultaneous rotation of rotor angles $\tht_j \to \tht_j + \tht$ by any fixed $\tht$ is a symmetry of $H_{\rm tot} = H_{\rm CM} + H$. It takes $\vf_0 \to \vf_0 + \tht$ but acts trivially on the relative angles $\vf_{1,2}$ so that it is a symmetry of $H_{\rm CM}$ but not $H$. The system also admits discrete symmetries. $H_{\rm tot}$ is symmetric under $G = S_3 \times \mathbb Z_2$ where $S_3 = \{ e, \tau_{12}, \tau_{23}, \tau_{31}, \sig_{123}, \sig_{132} \}$ is the group of permutations of the rotor angles and $\mathbb Z_2 = \{ e, \pi \}$ is the parity group with $\pi$ taking $\tht_j \to -\tht_j$ (see Table \ref{t:action-of-S3-Z2-on-angles}). These continue to be symmetries of the relative Hamiltonian $H$ with the action on relative angles given in Table \ref{t:action-of-S3-Z2-on-angles}. The CM angle $\vf_0$ is invariant under rotor permutations but reverses sign under parity. As a consequence, $H_{\rm CM}$ is also invariant under $G$. These discrete symmetries will play a crucial role in decomposing the spectrum of the Hamiltonian into irreducible representations of $G$.

\begin{table}
\begin{center}
\begin{tabular}{|c|c|c|c|}
\hline
g & acts on $(\tht_1, \tht_2, \tht_3)$ & acts on $(\vf_1, \vf_2)$ & acts on $(m,n)$\\
\hline
$e$ & $(\tht_1, \tht_2, \tht_3)$ & $(\vf_1, \vf_2)$ & $(m,n)$\\
$\pi$ & $(-\tht_1, -\tht_2, -\tht_3)$ & $(-\vf_1, -\vf_2)$ & $(-m,-n)$\\
$\tau_{12}$ & $(\tht_2, \tht_1, \tht_3)$ & $(-\vf_1, \vf_1 + \vf_2)$ & $(n-m, n)$\\
$\tau_{23}$ & $(\tht_1, \tht_3, \tht_2)$ & $(\vf_1 + \vf_2, -\vf_2)$ & $(m, m-n)$\\
$\tau_{31}$ & $(\tht_3, \tht_2, \tht_1)$ & $(-\vf_2, -\vf_1)$ & $(-n, -m)$\\
$\sigma_{123}$ & $(\tht_2, \tht_3, \tht_1)$ & $(\vf_2, -\vf_1 -\vf_2)$ & $(-n, m-n)$\\
$\sigma_{132}$ & $(\tht_3, \tht_1, \tht_2)$ & $(-\vf_1 -\vf_2, \vf_1)$ & $(n-m, -m)$\\
\hline
\end{tabular}
\end{center}
\caption{\small Action of the elements of the parity $\mZ_2 = \{ e, \pi \}$ and permutation $S_3 = \{ e, \tau, \sig \}$ groups on rotor angles $\tht_{1,2,3}$, relative angles $\vf_{1,2}$ and the Fourier basis element $(m,n) \equiv e^{i(m \vf_1 + n \vf_2)}/2\pi$. The action of ordered pairs of elements of $\mZ_2$ and $S_3$ is obtained by composition. Under this action, $\mZ_2 \times S_3$ is a symmetry of $H_{\rm tot}$ as well as of $H$. 
}
\label{t:action-of-S3-Z2-on-angles}
\end{table}

\section{Some approximate results}
\label{s:approx-results}

Here we discuss variational, perturbative and harmonic approximations that allow us to understand the spectrum of the relative Hamiltonian $\hat H$ in limiting cases. They also help to validate our numerical approach to diagonalizing the Hamiltonian in Section \ref{s:numerical-diagonalization}.

\subsection{Variational approach}

\begin{figure}
	\begin{subfigure}{0.60\columnwidth}
	\includegraphics[width = \textwidth]{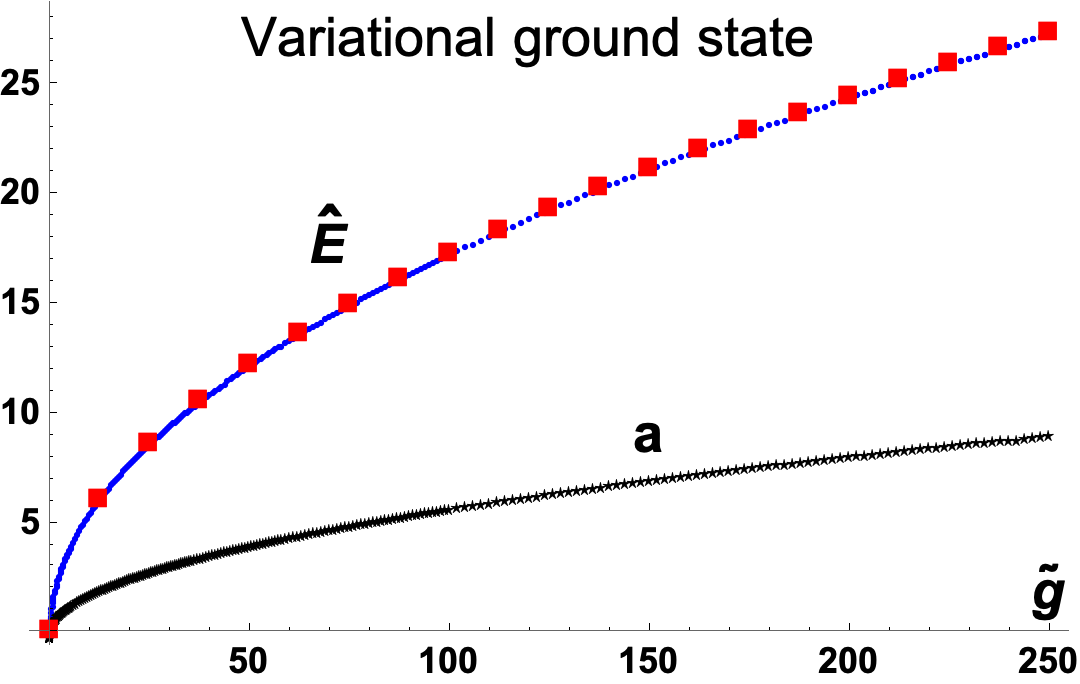}
	\caption{}
	\end{subfigure}
	\begin{subfigure}{0.38\columnwidth}
	\includegraphics[width = \textwidth]{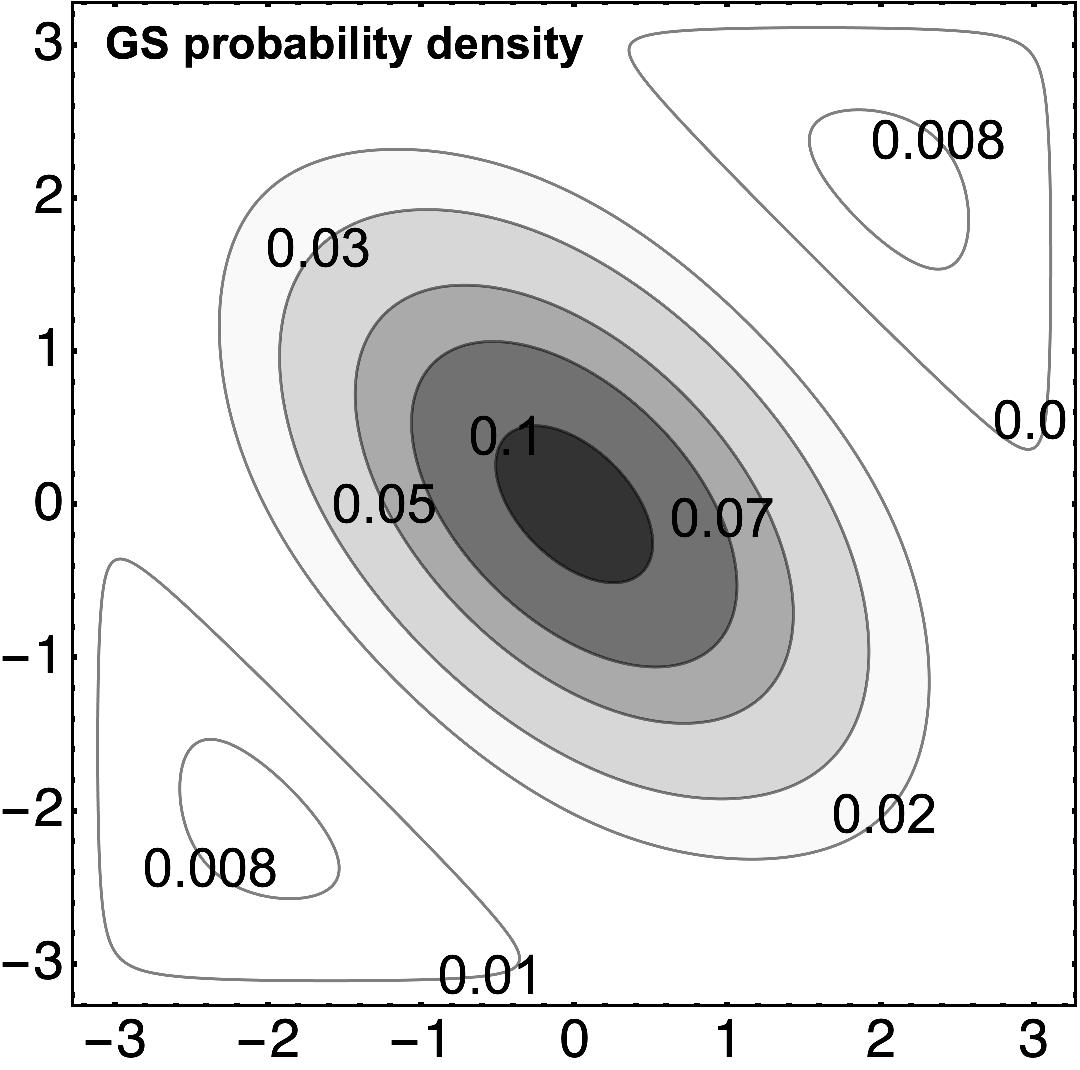}
	\caption{}
	\end{subfigure}
	\caption{\small (a) Variational ground state energies of $\hat H$ (blue dots) as a function of $\tl g$ based on the ansatz $\psi \propto e^{-aV}$ are very close to the numerical values (red squares). Optimal values of the variational parameter $a$ are also shown (black stars). (b) The variational ground state normalized probability density on the $-\pi \leq \vf_1, \vf_2 \leq \pi$ torus for the intermediate value $\tlg \approx 0.3$ where $a = 0.3$. The state is weakly localized around the minimum $(0,0)$ of the potential $V$ (\ref{e:Hamiltonian-total-CM-rel}).}
	\label{f:variational-E-and-a}
\end{figure}

We will obtain a variational upper bound on the ground state energy
	\beq
	\hat E_0 = \min_{\psi} \bra \psi | \hat H | \psi \ket /\bra \psi | \psi \ket
	\eeq
by minimizing over a class of wave functions obeying appropriate boundary conditions. For $\varpi = 1$, from (\ref{e:BCs-3torus}), the wave functions and their first partial derivatives in $\vf_{1,2}$ must be $2\pi$-periodic. A constant trial function $\psi$ leads to the crude upper bound $\hat E_0 \leq 3 \tl g$. A better ansatz with dimensionless variational parameter $a$ is given by $\psi(\vphi_1,\vphi_2) = e^{-aV}$, which automatically satisfies the BCs as $V(\vphi_1,\vphi_2)$ (\ref{e:Hamiltonian-total-CM-rel}) is periodic. The expectation values and optimal values of $a$ are found numerically for $10^{-3} \leq \tl g \leq 100$. The resulting variational estimates for $\hat E_0(\tl g)$ and optimal $a(\tl g)$ are plotted in Fig.~\ref{f:variational-E-and-a}. They compare favorably with the results of numerical diagonalization in Section \ref{s:numerical-diagonalization}. For instance, for $\tl g=1$, the variational estimate $\hat E_0^{\rm var} \approx 1.62$ for the optimal value $a \approx .68$ is quite close to the numerically obtained $\hat E_0 \approx 1.593$. By fitting the variational estimates over the above range of $\tl g$, we find the following asymptotic behaviors:
	\beqs
	\hat E^{\rm var}_0 &\approx& \begin{cases} 3 \tl g - 1.5 {\tl g}^2 & \text{as} \quad \tl g \to 0 \\
	\sqrt{3 \tlg} & \text{as} \quad \tl g \to \infty 
	\end{cases}  \quad \text{and}\cr
	a &\approx& \begin{cases} \tl g & \text{as} \quad \tl g \to 0 \\
	\sqrt{\tl g / e} & \text{as} \quad \tl g \to \infty. 
	\end{cases}
	\label{e:var-Ehat-vs-gtil-asymptotes}
	\eeqs
In the weak coupling limit ($\tl g \to 0$), this variational estimate matches the ground state energy obtained from second order perturbation theory in Section \ref{s:pert-theory}. Moreover, in this limit, $a \to 0$ and the optimal trial wave function approaches the constant ground state wave function of noninteracting rotors. On the other hand, in the semi-classical ($\tl g \to \infty$) limit, $a \to \infty$, and $\psi$ is concentrated at the minimum of $V$ with  $\tl E^{\rm var}_0 \sim \sqrt{3/\tlg} \to 0$, in agreement with the vanishing energy (in units of $g$) of the classical ground state \cite{gskhs-3rotor}.

\subsection{Perturbative approach}
\label{s:pert-theory}

\begin{figure}
	\centering
	\begin{tikzpicture}
	\coordinate (Origin)   at (0,0);
	\coordinate (XAxisMin) at (-1,0);
	\coordinate (XAxisMax) at (4,0);
	\coordinate (YAxisMin) at (0,-2);
	\coordinate (YAxisMax) at (0,4);
	\draw [thin,-latex] (XAxisMin) -- (XAxisMax) node[right] {$x$};
	\draw [thin,-latex] (YAxisMin) -- (YAxisMax) node[above] {$y$} ;
	\clip (-2.5,-2.5) rectangle (3.6cm,3.6cm); 
\begin{scope} 
\pgftransformcm{1}{0}{1/2}{sqrt(3)/2}{\pgfpoint{0cm}{0cm}}; 
\coordinate (Bone) at (0,2);
\coordinate (Btwo) at (2,-2);
\coordinate (1) at (2,0);
\coordinate (w) at (-2,2);
\draw[] (-4,-4) grid[step=2cm] (4,4);
\foreach \x in {-4,-3,...,4}{
  \foreach \y in {-4,-3,...,4}{
    \coordinate (Dot\x\y) at (2*\x,2*\y);
    \node[draw,circle,inner sep=2pt,fill] at (Dot\x\y) {};
  }
}
\draw [ultra thick,-latex,red,dashed] (Origin)
    -- ($(1)+(w)$) node [above]  {$-\om^2 \; (1,1)$};
\draw [ultra thick,-latex,red,dashed] (Origin)
    -- ($-1*(w)$) node [below]  {$-\om \; (0,-1)$};
\draw [ultra thick,-latex,red,dashed] (Origin)
    -- (1) node [below] {$1 \; (1,0)$};
 \draw [ultra thick,-latex,red,dashed] (Origin)
    -- ($-1*(1)$) node [below] {$\quad\;-1 \; (-1,0)$};
     \draw [ultra thick,-latex,red,dashed] (Origin)
    -- (w) coordinate (B3) node [above] {$\om \; (0,1)$};
     \draw [ultra thick,-latex,red,dashed] (Origin)
    -- ($-1*(1)-1*(w)$) node [below]  {$\om^2 \; (-1,-1)$};
\draw [ultra thick,-latex,blue] (Origin)
    -- ($(1)+2*(w)$) node [below right]  {$(1,2)$};    
\draw [ultra thick,-latex,blue] (Origin)
    -- ($(1)-1*(w)$) node [below]  {$(1,-1)$};  
\draw [ultra thick,-latex,blue] (Origin)
    -- ($2*(1)+(w)$) node [above]  {$(2,1)$};  
\end{scope} 
\begin{scope}
\pgftransformcm{1}{0}{-1/2}{sqrt(3)/2}{\pgfpoint{0cm}{0cm}}; 
\draw[] (-4,-4) grid[step=2cm] (4,4);
\end{scope}
	\end{tikzpicture}
	\caption{\small Eigenstates of kinetic energy $T$, represented as $(m, n)$ pairs on the $\mathbb{Z} + \om \mathbb{Z}$ lattice on the complex plane for $\om = e^{2\pi i/3}$. Rotations by elements of the cyclic group $C_6 = \bra e^{i \pi/3} \ket$ are symmetries of the lattice that preserve the distance $|m + n \om|$ from the origin (and hence preserve the eigenvalues of $T$). So the cardinality of each orbit must be divisible by $6$ implying that the degeneracies are multiples of 6. Dashed (red) and solid (blue) arrows point to two such orbits corresponding to energies $T = 1$ and $T=3$.}
	\label{f:triangular-lattice}
\end{figure}
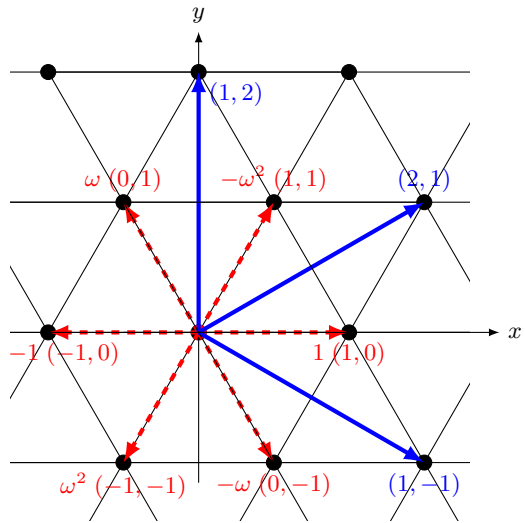

Here we use first order perturbation theory for the Hamiltonian $\hat H = T + \tlg V$ (\ref{dimless-H}) to estimate the energies of several low-lying states. The eigenvalues of $T$ are of the form $m^2 + n^2 - mn$ where $m$ and $n$ are integers corresponding to the Fourier eigenmodes $e_{m,n} = e^{i (m \vf_1 + n \vf_2)}/2\pi$. The first few eigenenergies $T$ along with their degeneracies $d$ are $T^d = 0^1$, $1^6$, $3^6$, $4^6$, $7^{12}$, $9^6$, $12^6$ and $13^{12}$. Aside from the ground state, the degeneracies must be multiples of six. To see this, notice that $ m^2 + n^2 -mn = |m + n \om|^2$ where $\om = e^{2\pi i/3}$ (or $e^{4\pi i/3}$) is a nontrivial cube root of unity. Hence, we may view the energy eigenvalue $m^2 + n^2 - mn$ for $m,n \in \mathbb{Z}$ as the square of the distance from the origin to the point $m + n\om$ lying on the triangular integral lattice spanned by $(1,\om)$ in the complex plane, as shown in Fig.~\ref{f:triangular-lattice}. The vertices of this lattice labelled $(m, n)$ represent the linearly independent eigenstates $e_{m, n}$ of the kinetic energy $T$. The cyclic group $C_6$ of rotations generated by $e^{i\pi/3}$ (sixth roots of unity) is a symmetry of the lattice that preserves the distance from the origin. Thus, the number of lattice points at a fixed distance from the origin must be a multiple of six. In fact, using further properties of the Unique Factorization Domain $\mathbb{Z} + \mathbb{Z} \om$, one may obtain an explicit formula for the degeneracies. Suppose the eigenenergy $T$ has the prime factorization $3^\g p_1^{\al_1} \cdots p_r^{\al_r} q_1^{\beta_1} \cdots q_s^{\beta_s}$, where $p_i \equiv 1 \; \mod 3$ and $q_i \equiv 2 \; \mod 3$ with $\beta_i$ even. Then the degeneracy of the eigenvalue $T$ is (see Appendix \ref{a:degeneracy-formula-unperturbed})
	\beq
	\text{degen}(T) = 6 (\al_1 + 1) (\al_2 + 1) \cdots (\al_r + 1).
	\label{e:degeneracy-free-rotor-levels}
	\eeq
If any of the $\beta_i$ is odd, then $T$ cannot arise as an eigenvalue and the `degeneracy' vanishes. For example, $T \in \{1, 3, 4 = 2^2, 9 = 3^2, 12 = 2^2 3 \}$ have no prime factors congruent to 1 modulo 3 and the $\beta$'s are even so that all these degeneracies are equal to 6. $T = 7$ and $T = 13$ are both primes $\equiv 1$ $(\mod 3)$ so that $r=1$, $\al_1 = 1$ and $\g = s = 0$ leading to a degeneracy of 12 for both levels. These multiplicities agree with the ones given at the beginning of this section.

The ground state $e_{0,0}$ of the noninteracting system $T$ is nondegenerate with $\hat E_0^{(0)} = 0$ and $\psi_0^{(0)} = 1/2\pi$. This implies that 
	\beq
	\hat E_0^{(1)} = \bra 1/2\pi \vert V \vert 1/2\pi \ket = 3.
	\eeq
Thus, to the first order, $\hat E_0 \approx \hat E_0^{(0)} + \tl g \hat E_0^{(1)} = 3 \tl g$. At second order, $\hat E_0^{(2)} = - 1.5$, so that $\hat E_0 \approx 3 \tl g - 1.5 \tl g^2$. For small $\tl g$, this agrees well with the numerically obtained ground state energy as well as our variational approximation (see (\ref{e:var-Ehat-vs-gtil-asymptotes}) and Fig.~\ref{f:variational-E-and-a}).

The first excited state of $T$ is 6-fold degenerate with energy $\hat E_1^{(0)} = 1$ and the eigenspace is spanned by $e^{\pm i \vf_1}/2\pi$, $e^{\pm i \vf_2}/2\pi$ and  $e^{\pm i (\vf_1+\vf_2)}/2\pi$. We find that the matrix representation of $V$ has eigenvalues $2$, $2.5$, $2.5$, $3.5$, $3.5$ and $4$. Thus, to first order in degenerate perturbation theory, we have
	\beqs
	&\hat E_1 = 1 + 2 \tl g, \quad \hat E_{2,3} = 1+ 2.5 \tl g \cr
	&\hat E_{4,5} = 1+ 3.5 \tl g \quad \text{and} \quad \hat E_6 = 1 + 4 \tl g.
	\eeqs
We find that this agrees well with the low-lying spectrum derived numerically for small $\tlg \lesssim 0.15$. For the second and third excited states of $T$, first order perturbation theory does not break the 6-fold degeneracies, and we get $\hat E_{7-12} = 3 + 3 \tl g$ and $\hat E_{13-18} = 4 + 3 \tl g$. However, the 12-fold degenerate $4^{\rm th}$ excited state of $T$ splits into two 6-fold degenerate levels $\hat E_{19-24} = 7 + 2.5 \tl g$ and $\hat E_{25-30} = 7 + 3.5 \tl g$.


\subsection{Harmonic approximation around ground state} 
\label{s:harmonic-approx}

In Jacobi-like coordinates $\vf_\pm = \vf_1 \pm \vf_2$, the kinetic energy operator (\ref{e:Hamiltonian-total-CM-rel}) becomes diagonal: 
	\beqs
	\hat H &=& - \left( \pdr^2_{\vf_+} + 3 \pdr^2_{\vf_-} \right) - \tl g \bigg[ \cos \fr{\vf_+
+ \vf_-}{2} \cr
&& \quad + \cos \fr{\vf_+ - \vf_-}{2} + \cos \vf_+ -3 \bigg]. 
	\eeqs
In the harmonic approximation around the classical ground state, $\vf_\pm$ are small and the Hamiltonian
	\beq
	\hat H^{\rm harm} = - \left( \pdr^2_{\vf_+} + 3 \pdr^2_{\vf_-} \right) + \fr{\tl g }{4}(3\vf_+^2 + \vf_-^2)
	\label{e:harm-approx-hamil}
	\eeq
splits into a pair of uncoupled linear oscillators so that
	\beq
	\hat E^{\rm harm}_{m,n} = \sqrt{3\tl g}  (m + n + 1) \quad \text{for} \quad m,n = 0,1, 2, \cdots.
	\label{e:harm-spectrum}
	\eeq
Pleasantly, the $S_3 \times \mathbb{Z}_2$ symmetry of the three-rotor system survives the harmonic approximation. The action of group elements on $\vphi_\pm$ deduced from Table \ref{t:action-of-S3-Z2-on-angles} are given in Table \ref{t:action-of-G-on-phipm}. It turns out that $\hat H^{\rm harm}$ (\ref{e:harm-approx-hamil}) is invariant under this group action as we now argue. The kinetic energy is unaffected by the approximation. To see the invariance of the potential, we note that up to an additive constant, it is proportional to $\vf_1^2 + \vf_2^2 + (\vf_1 + \vf_2)^2$ and observe that elements of $S_3$ permute the three terms while parity just reverses the signs of $\vf_{1,2}$. The implementation of this symmetry in the harmonic approximation will be relevant to the purification of the spectrum to be discussed in Section \ref{s:harmonic-approx-purification-spacing}.

\begin{figure}	
	\includegraphics[width=8cm]{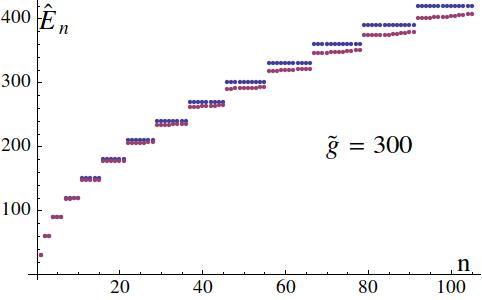}
	\caption{\small Lowest 100 numerically obtained eigenvalues of $\hat H$ (red, lower) follow a pattern similar to that of the 2d harmonic oscillator (blue, higher) provided $\tl g$ is large.}
	\label{f:SHO-comparision}
\end{figure}

For large $\tl g \approx 300$, we find that energies of low-lying states are well captured by the harmonic approximation as seen in Fig.~\ref{f:SHO-comparision}. Although degeneracies of the 2d harmonic oscillator are somewhat broken, the multiplet structure ($n^{\rm th}$ energy level being $n$-fold degenerate) is approximately retained for the lowest few levels. This is to be expected because for large $\tl g$ (stiff spring), the low-lying states are strongly localized around the classical ground state where the harmonic part of the potential dominates. As $\tl g$ is increased, a larger number of low-lying levels (roughly when $m+n+1 \ll \sqrt{\tlg/ 3}$) follow the harmonic approximation. This observation suggests that we should omit these low-lying states when comparing spectral statistics with Poisson or Wigner-Dyson spacing distributions as the spacing of an oscillator is quite different from the latter ones, see Fig.~\ref{f:spacing-distrib}.

\begin{table}
\begin{center}
\begin{tabular}{|c|c|c|c|}
\hline
g & action on $(\vf_+, \vf_-)$  \\
\hline
$e$ & $(\vf_+, \vf_-)$ \\
$\pi$ & $(-\vf_+, -\vf_-)$ \\
$\tau_{12}$ & $((\vf_+ - \vf_-)/2, -(3\vf_+ + \vf_-)/2$ \\
$\tau_{23}$ & ($(\vf_+ + \vf_-)/2, (3\vf_+ - \vf_-)/2$) \\
$\tau_{31}$ & $(-\vf_+, \vf_-)$ \\
$\sigma_{123}$ & $(-(\vf_++\vf_-)/2, (3\vf_+ -\vf_-)/2)$ \\
$\sigma_{132}$ & $(-(\vf_+ - \vf_-)/2, -(3\vf_+ + \vf_-)/2)$ \\
\hline
\end{tabular}
\end{center}
\caption{\small Action of the elements of the parity $\mZ_2 = \{ e, \pi \}$ and permutation $S_3 = \{ e, \tau, \sig \}$ groups on Jacobi angles $\vf_\pm$.}
\label{t:action-of-G-on-phipm}
\end{table}

\section{Numerical diagonalization and decomposition of the spectrum}
\label{s:numerical-diagonalization}

In this section, we first describe a scheme to determine the spectrum of the relative Hamiltonian by numerical diagonalization. The cumulative distribution function of this `total spectrum' is then compared with a Weyl-type semiclassical asymptotic formula. We then address the question of decomposing the total spectrum into various irreducible representations of the discrete symmetry group $G = S_3 \times \mathbb{Z}_2$. Finally, an explicit basis for $G$-invariant states is constructed to compute the spectrum in the identity representation. 

\subsection{Total spectrum via Fourier basis}

As mentioned in \ref{s:SOV-BC}, for zero center of mass angular momentum ($\ell = 0$), the wave function $\psi$ must be $2\pi$-periodic in both $\vf_1$ and $\vf_2$. Such wave functions may be expanded in an orthonormal Fourier basis 
	\beq
	\bra \vf_1 ,\vf_2| m, n \ket = e_{m,n} = e^{i (m \vf_1 + n \vf_2)}/2\pi \equiv (m,n).
	\label{e:fourier-basis}
	\eeq
Noting that the potential $V$ in (\ref{e:Hamiltonian-total-CM-rel}) can be rewritten as
	\beqs
	\label{e:hamiltonian-exp-pot}
	V &=&  - \ov{2} \left( e^{i \vf_1} + e^{-i \vf_1} + e^{i \vf_2} + e^{-i \vf_2} \right) \cr 
	  &&- \ov{2} \left( e^{i (\vf_1 + \vf_2)} + e^{-i (\vf_1 + \vf_2)}  \right) + 3,
	\eeqs
the matrix elements of $\hat H$ are
	\beqs
	\hat H_{a,b;m,n} &=& \bra a,b | \hat H | m,n \ket \cr
	&=& \delta^a_m \delta^b_n \left(3 \tlg + [m^2 + n^2 - m n] \right) 
	\cr
	&& - \fr{\tlg}{2} \left[ \delta^b_n (\delta^a_{m+1} + \delta^a_{m-1})+ \del^a_m (\del^b_{n+1} + \delta^b_{n-1}) \right] \cr
	&&- \fr{\tlg}{2} \left[\delta^a_{m+1}\delta^b_{n+1} + \delta^a_{m-1}\delta^b_{n-1}  \right].
	\eeqs
To find the eigenvalues of $\hat H$, we proceed as follows. First, we truncate the Fourier basis $e_{m,n}$ to $0 \leq |m|, |n| \leq m_{\rm max}$ corresponding to the $(2 m_{\rm max} + 1) \times (2 m_{\rm max} + 1)$ integer lattice on the plane centered at the origin. To get an ordered basis, we traverse this lattice columnwise upwards starting from the south-west  corner. This leads us to an ordered orthonormal Fourier basis $e_i$ labeled by the single index $i(m,n) = (2 m_{\rm max} + 1) m + n + 2m_{\rm max}^2 + 2 m_{\rm max} + 1$ with $1 \leq i \leq (2 m_{\rm max} + 1)^2$. The matrix elements $\hat H_{ij} = \bra e_i | \hat H | e_j \ket$  are computed in this basis resulting in a truncated Hamiltonian matrix of dimension $N = (2 m_{\rm max} + 1)^2$ that we diagonalize numerically. However, the larger eigenvalues suffer from truncation errors. By requiring that the eigenvalues do not change appreciably when $N$ is increased, we isolate the reliably computed portion of the spectrum. 

\subsubsection{Density of states and Weyl's law}

\begin{figure}
	\includegraphics[width=8.5cm]{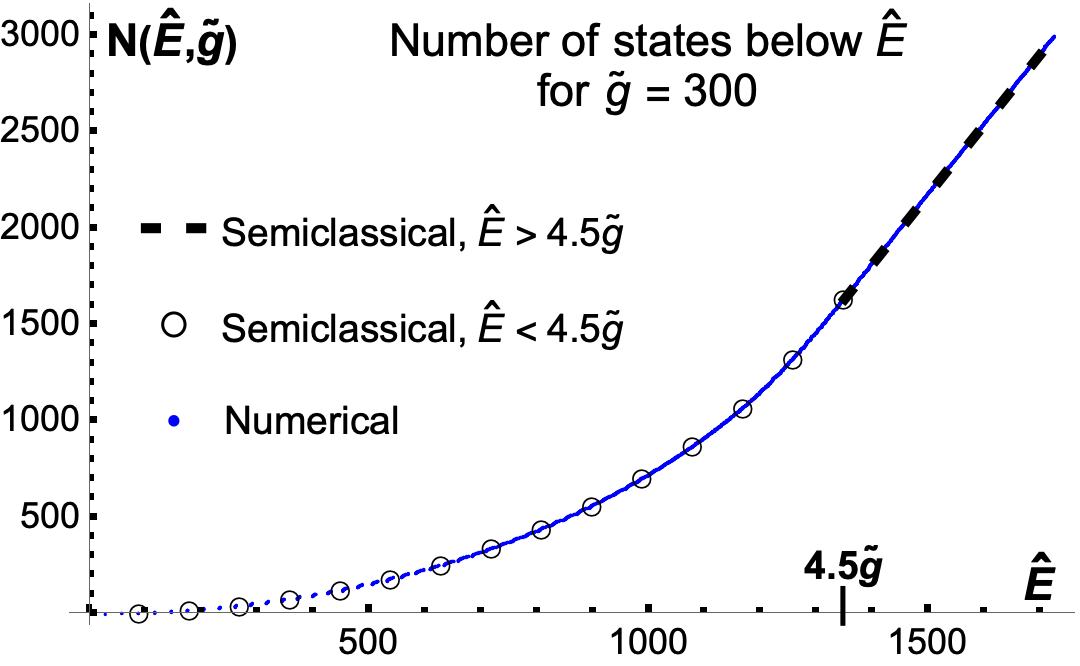}
	\caption{\small Numerically computed number of energy levels $N(\hat E, \tl g)$ below energy $\hat E$ for $\tl g = 300$ (blue dots) agrees well with semiclassical estimate of Eq. (\ref{e:weyl-formula-phi12-integ}) (black dashes for $\hat E > 4.5 \tl g$ and black circles for $\hat E < 4.5 \tl g$). Asymptotically, $N \sim (2\pi/\sqrt{3})(\hat E - 3 \tlg)$.}
	\label{f:weyl-cdf-compare}
\end{figure}

A version of Weyl's semiclassical law \cite{weyl-1912} for systems with a potential says that the asymptotic number $N(E, \hbar)$ of energy eigenvalues below $E$ is given by
	\beq
	N(E, \hbar) \sim \ov{(2 \pi \hbar)^2} \int_{H < E} d \vf_1 d p_1 d \vf_2 d p_2.
	\eeq
The RHS is simply the energetically-allowed phase space volume in units of $h^2 = (2 \pi \hbar)^2$. Since $N(E,\hbar)$ is dimensionless, it can depend on $E$ and $\hbar$ only via the dimensionless energy $\hat E = E/(\hbar^2mr^2)$ and coupling $\tl g = gmr^2/\hbar^2$. Introducing the dimensionless angular momenta $\tl p_i = p_i/\sqrt{g m r^2}$ allows us to write
	\beq
	N(\hat E, \tl g) \sim \frac{\tl g}{(2\pi)^2} \int_{\hat H \leq \hat E} d\vf_1 d \tl p_1 d \vf_2 d \tl p_2.
	\label{e:weyl-law-dimless}
	\eeq
Here, from (\ref{e:Hamiltonian-total-CM-rel}),
	\beqs
	\hat H &=& T + \tl g V = \tl g [ \tl p_1^2 -\tl p_1 \tl p_2 + \tl p_2^2 \cr 
	&& + 3 -\cos \vf_1 - \cos \vf_2 - \cos(\vf_1 + \vf_2) ].
	\eeqs
Formula (\ref{e:weyl-law-dimless}) is expected to hold in a semiclassical limit where $\hat E \gg 1$ and/or $\tl g \gg 1$. To evaluate (\ref{e:weyl-law-dimless}), the angular momentum integral is performed first for fixed $(\vf_1, \vf_2)$ followed by an integral over the energetically allowed Hill region of the configuration torus. For each point $(\vf_1, \vf_2)$, the energetically allowed region $R_{\vf_1, \vf_2}$ in angular momentum space is given by $T \leq \hat E - \tl g V$. Defining $K(\vf_1, \vf_2) = \hat E/\tlg - V(\vf_1, \vf_2)$, the boundary of this region is the ellipse
	\beq
	\tl p_1^2 - \tl p_1 \tl p_2 + \tl p_2^2 = K(\vf_1, \vf_2).
	\eeq
It may be brought to the standard form $\tl p_+^2/(2K) + \tl p_-^2/(2K/3) = 1$ upon defining $\tl p_\pm = (\tl p_1 \pm \tl p_2)/\sqrt{2}$, whence it encloses an area $2\pi K/\sqrt{3}$. Consequently,
	\beqs
	N &\sim& \frac{\tl g}{(2\pi)^2} \iint_{K(\vf_1, \vf_2) \geq 0} d \vf_1 d \vf_2 \int_{R_{\vf_1, \vf_2}} d \tl p_1 d \tl p_2 \cr
	&=& \frac{\tl g}{(2\pi)^2} \iint_{K \geq 0} d \vf_1 d \vf_2 \: (2 \pi/\sqrt{3}) K(\vf_1, \vf_2).\qquad
	\label{e:weyl-formula-phi12-integ}
	\eeqs
The angular integrals simplify if the rotors can explore the whole of the configuration torus. This happens if $\hat E > 4.5 \tl g$ because $T \geq 0$ and $0 \leq V \leq 4.5$ leading to $N \sim (2\pi/\sqrt{3})(\hat E - 3 \tlg)$. When $\hat E < 4.5 \tl g$, the integral (\ref{e:weyl-formula-phi12-integ}) extends only over the energetically allowed region $V < \hat E/\tlg$ of the $\vf_1-\vf_2$ torus and is evaluated numerically. Fig.~\ref{f:weyl-cdf-compare} shows that these semiclassical estimates are in good agreement with the numerically obtained cumulative distribution of levels for $\tl g = 300$.

\subsection{Purification of the spectrum}
\label{s:purify-spectrum}

The Hamiltonian possesses the discrete symmetry $G = S_3 \times \mZ_2$ (see \ref{s:sym-of-3rotor}). Thus the Hilbert space $\cal H$ spanned by the Fourier modes $e_{m,n}$ may be written as a direct sum of subspaces that carry irreducible representations of $G$. The Hamiltonian is block diagonal with respect to this decomposition. To examine spectral statistics and its universality, we must `desymmetrize' or `purify' the spectrum by isolating the energy levels corresponding to each irreducible block since the universal Poisson and Wigner-Dyson spacing distributions expected in integrable and chaotic systems do not incorporate any such discrete symmetries \cite{bohigas-giannoni-schmit,berry-tabor}.

\subsubsection{Unitary irreducible representations of \texorpdfstring{$S_3 \times \mZ_2$}{S-3 x Z-2}} 

The group $S_3$ (see Table \ref{t:action-of-S3-Z2-on-angles}) has two 1d irreducible representations (i) {\it identity} or {\it trivial}: $\rho_i(g) = 1$ $\forall g \in S_3$, (ii) {\it sign}: $\rho_s(g) = -1$ for transpositions and $\rho_s(g) = 1$ for the remaining permutations and (iii) a 2d unitary representation coming from the dihedral symmetries of an equilateral triangle centered at the origin with vertices labeled counterclockwise and with vertex  2 on the upper vertical axis: \small
	\beqs
	\rho_d(e) = \begin{smmat} 1 & 0 \\ 0 & 1 \end{smmat},
	&& \quad 
	\rho_d(\tau_{12}) = \half \begin{smmat} 1 & \sqrt{3} \\ \sqrt{3} & - 1 \end{smmat}, \cr
	\rho_d(\tau_{23}) = \half \begin{smmat} 1 & -\sqrt{3} \\ -\sqrt{3} & -1 \end{smmat}, && \quad
	\rho_d(\tau_{31}) = \begin{smmat} -1 & 0 \\ 0 & 1 \end{smmat}, \cr
	\rho_d(\sig_{123}) = \half \begin{smmat} -1 & -\sqrt{3} \\ \sqrt{3} & - 1 \end{smmat}, && \quad
	\rho_d(\sig_{132}) = \half \begin{smmat} -1 & \sqrt{3} \\ -\sqrt{3} & - 1 \end{smmat}. \quad
	\eeqs
\normalsize
On the other hand, $\mZ_2$ has only two irreducible representations, both of dimension one: the identity representation $\varrho_i(g) = 1$ for all $g \in \mathbb{Z}_2$ and the sign representation: $\varrho_s(e) = 1$ and $\varrho_s(\pi) = -1$. Since the tensor product of irreducible representations is an irreducible representation of the Cartesian product (see Section 2.6 of \cite{sternberg-group-theory}), we get six unitary irreducible representations of $S_3 \times \mathbb Z_2$:
	\beqs
	&&\text{(i) }\; \rho_{ii}(g_1, g_2) = \rho_{i} (g_1) \otimes \varrho_i (g_2),\cr
	&&\text{(ii) }\; \rho_{is}(g_1, g_2) = \rho_i (g_1) \otimes \varrho_s (g_2),\cr
	&&\text{(iii) }\; \rho_{si}(g_1, g_2) = \rho_s (g_1) \otimes \varrho_i (g_2),\cr
	&&\text{(iv) }\; \rho_{ss}(g_1, g_2) = \rho_s (g_1) \otimes \varrho_s (g_2),\cr
	&&\text{(v) }\; \rho_{di}(g_1, g_2) = \rho_d (g_1) \otimes \varrho_i (g_2) \quad \text{and} \cr
	&&\text{(vi) }\; \rho_{ds}(g_1, g_2) = \rho_d (g_1) \otimes \varrho_s (g_2)
	\eeqs
where $g_1 \in S_3$ and $g_2 \in \mZ_2$. These representations have dimensions $1,1,1,1,2,2$. The sum of the squares of the dimensions is 12, which is the same as the cardinality of $S_3 \times \mathbb{Z}_2$. Thus, by the dimension formula \cite{sternberg-group-theory}, these exhaust all the irreducible representations of $S_3 \times \mathbb{Z}_2$.

\subsubsection{Basis for the identity representation}

In what follows, for simplicity, we will restrict attention to the identity representation $\rho_{ii}$. To find the purified spectrum in this case, we need to diagonalize $H$ in the invariant subspace of $\cal H$ corresponding to $\rho_{ii}$. To this end, we seek a basis $s_{m,n}(\vf_1, \vf_2)$ for the identity representation. A simple way of obtaining such a basis is by `averaging'. Given a Fourier basis element $e_{m, n}(\vf_1, \vf_2)$, we find its orbit under the action of $S_3 \times \mZ_2$ (see Table \ref{t:action-of-S3-Z2-on-angles}) and define $s_{m,n}$ to be the (normalized) sum:
	\beqs
	 s_{m,n}  &=& c_{m, n}^{-1} ( e_{ m, n} + e_{n - m, n} + e_{m, m-n} + e_{-n, -m} \cr
	 &+& e_{-n, m-n} + e_{n-m, -m} + e_{-m, -n} + e_{m-n, -n}\cr
	 &+& e_{-m, n-m} + e_{n, m} + e_{n, n-m} + e_{m-n, m})
	\label{e:triv-basis-elt}
	\eeqs
where $c_{m, n}$ are normalization constants defined in (\ref{e:normalization-constants}). To enumerate the independent basis elements we need to restrict to a single $(m,n)$ pair from each orbit. In Appendix \ref{a:orth-basis}, we show that this is achieved by restricting to 
	\beq
	m \ge 0 \quad \text{and} \quad 0 \le n \le \lfloor m/2 \rfloor,
	\label{e:basis-label-range}
	\eeq
where $\lfloor \cdot \rfloor$ denotes the greatest integer part. We order them in increasing order of $m$ and in decreasing order of $n$ for fixed $m$: $s_{0,0}$, $s_{1,0}$, $s_{2,1}$, $s_{2,0}$, $s_{3,1}, \cdots$ as indicated in Fig~\ref{f:wedge-basis-identity-representation}.  It is noteworthy that all the Fourier basis states $e_{p,q}$ that appear in the formula (\ref{e:triv-basis-elt}) for a particular $s_{m,n}$ are eigenvectors of kinetic energy $T = \hat H(\tl g = 0)$ with the same eigenvalue. Thus each $s_{m,n}$ is an eigenstate of the kinetic energy: $T s_{m,n} = (m^2 + n^2 - mn) s_{m,n}$. Note that the above order does not correspond to increasing kinetic energy although $T$ does increase with decreasing $n$ for fixed $m$. Furthermore, each level of $T$ could contribute more than one symmetrized basis state $s_{m,n}$. For instance, $s_{7,0}$ and $s_{8,3}$ have the same kinetic energy $49$.

By virtue of the orthogonality of $e_{m,n}$, the $s_{m,n}$ are also orthogonal. If the normalization factors are chosen as
	\beq
	c_{m,n} = \begin{cases}
	12 & \text{if} \quad m = n = 0, \\
	2 \sqrt 6 & \text{if} \quad m \neq n = 0, \\
	2 \sqrt 6 & \text{if} \quad n = m/2 \neq 0 \quad \text{and} \\
	2 \sqrt 3 & \text{otherwise,}
	\end{cases}
	\label{e:normalization-constants}
	\eeq
then $s_{m,n}$ subject to (\ref{e:basis-label-range}) furnish an orthonormal basis for the identity representation:
	\beq	
	\langle  s_{a,b}| s_{m,n} \rangle = \del^a_m \del^b_n.
	\eeq

\subsubsection{Matrix elements of \texorpdfstring{$\hat H$}{H-hat} in \texorpdfstring{$s_{m,n}$}{s-mn} basis}

The matrix elements of $\hat H = T + \tlg V$ in the $s_{m,n}$ basis for the identity representation are \small
	\beqs
	T_{a,b;m,n} &=& (m^2 + n^2 - mn) \del^a_m \del^b_n \quad \text{and} \cr
	V_{a,b;m,n} &=& 3 \del^a_m \del^b_n - \fr{c^{-1}_{m,n}}{2} \big( c_{m+1,n} \del^a_{m+1} \del^b_n + c_{m,n-1} \del^a_m \del^b_{n-1} \cr
	&& + c_{m-1,n-1} \del^a_{m-1} \del^b_{n-1} + c_{m-1,n} \del^a_{m-1} \del^b_{n} \cr
	&& + c_{m,n+1} \del^a_{m} \del^b_{n+1} + c_{m+1,n+1} \del^a_{m+1} \del^b_{n+1} \big).
	\eeqs \normalsize
In this formula, if $(i,j)$ in $c_{i,j} \del^a_i \del^b_j$ does not satisfy (\ref{e:basis-label-range}), then it is to be replaced by the appropriate orbit representative that does satisfy (\ref{e:basis-label-range}). In other words, we replace 
	\beqs
	&&(0,\pm 1) \to (1,0), \quad (-1,j) \to (1,0), \quad (i,-1) \to (i+1,1)  \cr
	&& \text{and} \;\; (i,\lfloor i/2 \rfloor +1) \to (i, i - \lfloor i/2 \rfloor -1) \;\; \text{when} \;\; i > 0. \qquad
	\eeqs
Taking these replacements into account, we get
	\begin{widetext}
	\small
	\beqs
	V_{a,b;m,n} = 3 \del^a_m \del^b_n - \begin{cases}
	3 (c_{1,0}/c_{0,0}) \del^a_{1} \del^b_{0} & \text{if} \;\; m = n = 0, \\
	\fr{c^{-1}_{1,0}}{2} ( c_{2,0} \del^a_{2} \del^b_0 + 	2 c_{2,1} \del^a_2 \del^b_{1} 
	+ 2 c_{1,0} \del^a_{1} \del^b_{0} + c_{0,0} \del^a_{0} \del^b_{0} ) & \text{if} \;\; m = 1, n = 0, \\
	\fr{c^{-1}_{m,0}}{2} ( c_{m+1,0} \del^a_{m+1} \del^b_0 + 2 c_{m+1,1} \del^a_{m+1} \del^b_{1} + c_{m-1,0} \del^a_{m-1} \del^b_{0} +
	 2 c_{m,1} \del^a_{m} \del^b_{1} ) & \text{if} \;\;  m >1, n = 0  \\
	\fr{c^{-1}_{m,n}}{2} ( c_{m+1,n} \del^a_{m+1} \del^b_n + c_{m,n-1} \del^a_m \del^b_{n-1} +  c_{m-1,n-1} \del^a_{m-1} \del^b_{n-1} \cr + c_{m-1,m-n-1} \del^a_{m-1} \del^b_{m-n-1} + c_{m,m-n-1} \del^a_{m} \del^b_{m-n-1} + c_{m+1,m-n} \del^a_{m+1} \del^b_{m-n} ) & \text{if} \;\; \lfloor m/2 \rfloor = n \neq 0 \;\; \text{and} \\
	\fr{c^{-1}_{m,n}}{2} ( c_{m+1,n} \del^a_{m+1} \del^b_n + c_{m,n-1} \del^a_m \del^b_{n-1} + c_{m-1,n-1} \del^a_{m-1} \del^b_{n-1} \cr + c_{m-1,n} \del^a_{m-1} \del^b_{n} + c_{m,n+1} \del^a_{m} \del^b_{n+1} + c_{m+1,n+1} \del^a_{m+1} \del^b_{n+1} ) & \text{if} \;\; m \ge 4, 0 < n< \lfloor m/2 \rfloor.
	\end{cases}
	\label{e:mat-el-V-triv-rep}
	\eeqs
	\end{widetext}


\subsubsection{Spectrum in the identity representation}

For numerical calculations, we limit $0 \le m \le m_{\rm max}$. We define a single index $i(m,n) = \sum_{j=0}^{m} (\lfloor j/2 \rfloor + 1) - n$ to label the basis elements in the identity representation. The resulting truncated Hamiltonian matrix $\hat H$ in this basis is of dimension $( \lfloor m_{\rm max}/2 \rfloor + 1)(\lfloor m_{\rm max}/2 \rfloor + 2)$ when $m_{\rm max}$ is odd. To avoid truncation errors and ensure convergence, we stipulate that the fractional difference in energy eigenvalues be less than $10^{-10}$ when $m_{max}$ is increased from $339$ to $349$ corresponding to matrix size increasing from $28,000$ to $30,000$. This allows us to reliably determine the lowest $\approx 12,000$ eigenvalues.

\section{Spacing distributions}
\label{s:spacing-distrib}

Nearest neighbor energy level spacing distributions are a standard way of quantifying quantum manifestations of regular or chaotic dynamics \cite{haake, stockmann}. For generic integrable and chaotic systems, the spacing distributions of desymmetrized or purified levels are expected to display universal Poisson and Wigner-Dyson statistics \cite{bohigas-giannoni-schmit,berry-tabor}. To extract this universal behavior, we `unfold' the purified spectrum to have, on average, unit level spacing \cite{haake,gsk-vishnu-screwon}. Denoting energy levels by $E_i$ for $i = 1,2,3, \ldots, N$, the unfolded levels are defined as $\xi_i = \xi(E_i)$ where $\xi(E)$ is the polynomial (say of degree 7) that best fits the cumulative distribution function $n(E) = \sum_{i = 1}^N \tht(E - E_i)$ where $\tht$ is the right-continuous Heaviside step function. 

However, the classical three-rotor problem is neither fully chaotic nor integrable but shows order-chaos-order behavior with increasing energy. To interpret the spacing distributions, we partition the spectrum into energy windows where the classical dynamics is regular, mixed or chaotic and do the unfolding separately in each window. In the semiclassical regime of large $\tl g \gtrsim 1000$, this partitioning allows us to see striking quantum signatures of the regular to chaotic and chaotic to regular transitions. 

It is noteworthy that the lowest-lying energy levels do not follow Poisson statistics even in the semiclassical regime. This is to be expected from Berry and Tabor's observation \cite{berry-tabor} that the spectrum of a harmonic oscillator does not obey Poisson statistics. In fact, we find that the harmonic approximation of Sections \ref{s:harmonic-approx} and \ref{s:harmonic-approx-purification-spacing} is able to capture the spacing distribution in this regime. Thus the low energy regular regime shows a transition from oscillator-like to Poisson statistics.

On the other hand, when $\tlg$ is relatively small (e.g. $\tlg = 2,10$) semiclassical expectations are not a useful guide and the order-chaos transition is barely visible due to a paucity of levels with $\hat E < 6 \tlg$. However, as shown in Section \ref{s:low-g-free-rotor}, we find that the noninteracting rotor system provides a good approximation, especially at relatively high energies.

\subsection{Semiclassical regime of large \texorpdfstring{$\tlg$}{g-tilde}}
\label{s:semiclass-regime-large-g}

In the classical three rotor problem \cite{gskhs-3rotor}, we have regular behavior for $E \lesssim 3.5g$, a transition to chaos for $3.5g \lesssim E \lesssim 5g$ followed by a globally chaotic band $5.33g \lesssim E \lesssim 5.6g$. There is then a gradual return to regularity as $E \to \infty$. Thus, in the semi-classical limit of large $\tlg$, we expect to see quantum signatures of this behavior in spacing distributions in the corresponding energy windows. Working in the identity representation, we partition the energy spectrum and investigate the nearest-neighbor spacings of unfolded levels in various $\hat E$ windows. One expects to find an exponential spacing distribution $P_{\rm e}(s) = e^{-s}$ when the classical behavior is close to integrable and the Wigner surmise $P_{\rm w}(s) \approx (\pi s/2) e^{- \pi s^2/4}$ for classically chaotic behavior \cite{berry-tabor, haake}. The latter corresponds to the Gaussian Orthogonal Ensemble of random matrices that is appropriate to a real symmetric and time-reversal invariant Hamiltonian \cite{mehta}. When the classical dynamics is mixed (islands of regular behavior interspersed with chaotic regions), we expect to find a spacing distribution that may be crudely modeled by a Brody distribution \cite{brody}
	\beqs
	P(s; \nu) &=& (1+\nu) f(\nu) s^\nu e^{-f(\nu) s^{1+\nu}} \cr
	\text{where} \;\; f(\nu) &=& \G\left( (2+\nu)/(1+\nu) \right)^{1+\nu} 
	\eeqs
that interpolates between exponential ($\nu = 0$) and Wigner ($\nu = 1$). 

\begin{figure}
\captionsetup[subfigure]{font=footnotesize,labelfont=footnotesize}
	\begin{subfigure}{0.49\columnwidth}
	\includegraphics[width = \textwidth]{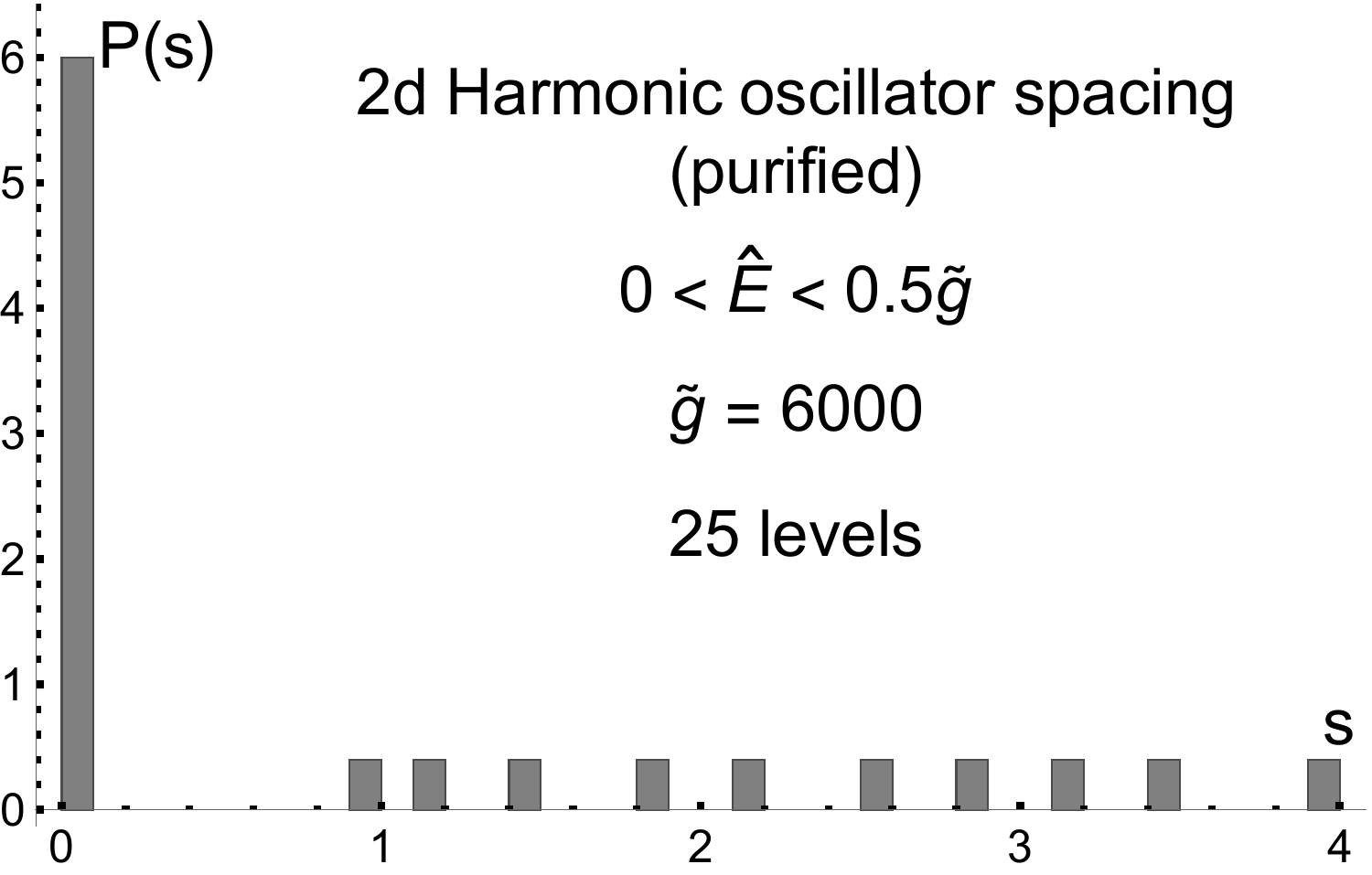}
	\caption{\small exact oscillator spacing}
	\label{f:g-6k-e-0-pt5g-harmonic}
	\end{subfigure}
	\begin{subfigure}{0.49\columnwidth}
	\includegraphics[width = \textwidth]{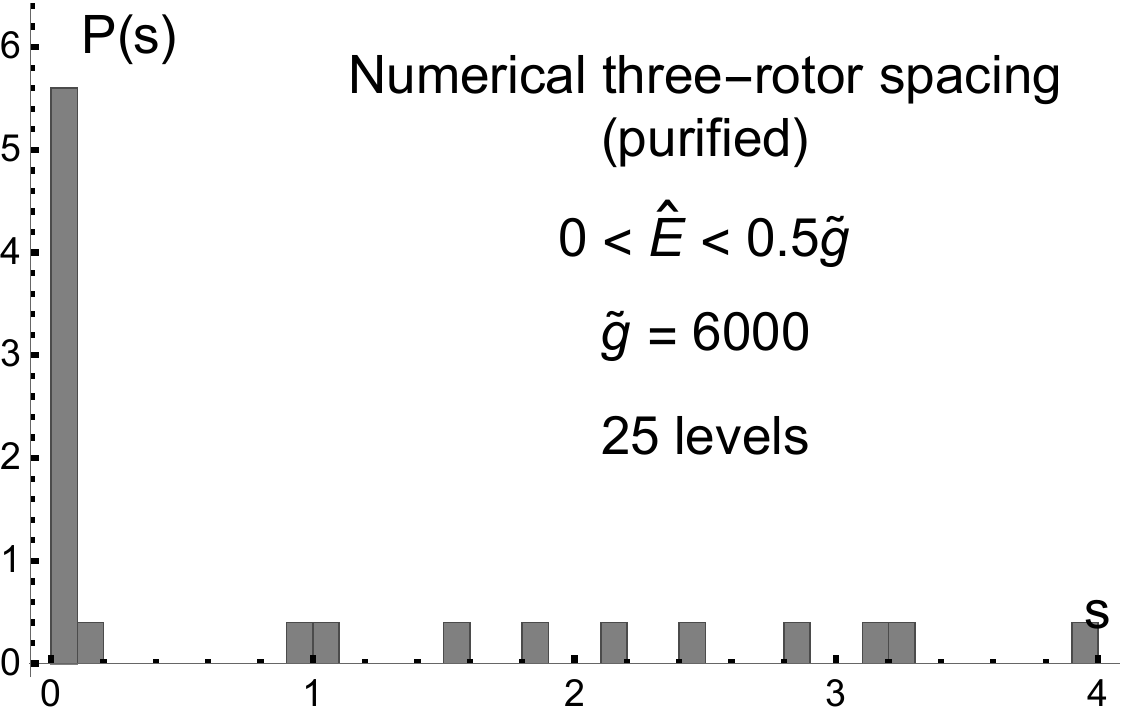}
	\caption{\small oscillator-like}
	\label{f:g-6k-e-0-pt5g}
	\end{subfigure}
	\begin{subfigure}{0.49\columnwidth}
	\includegraphics[width = \textwidth]{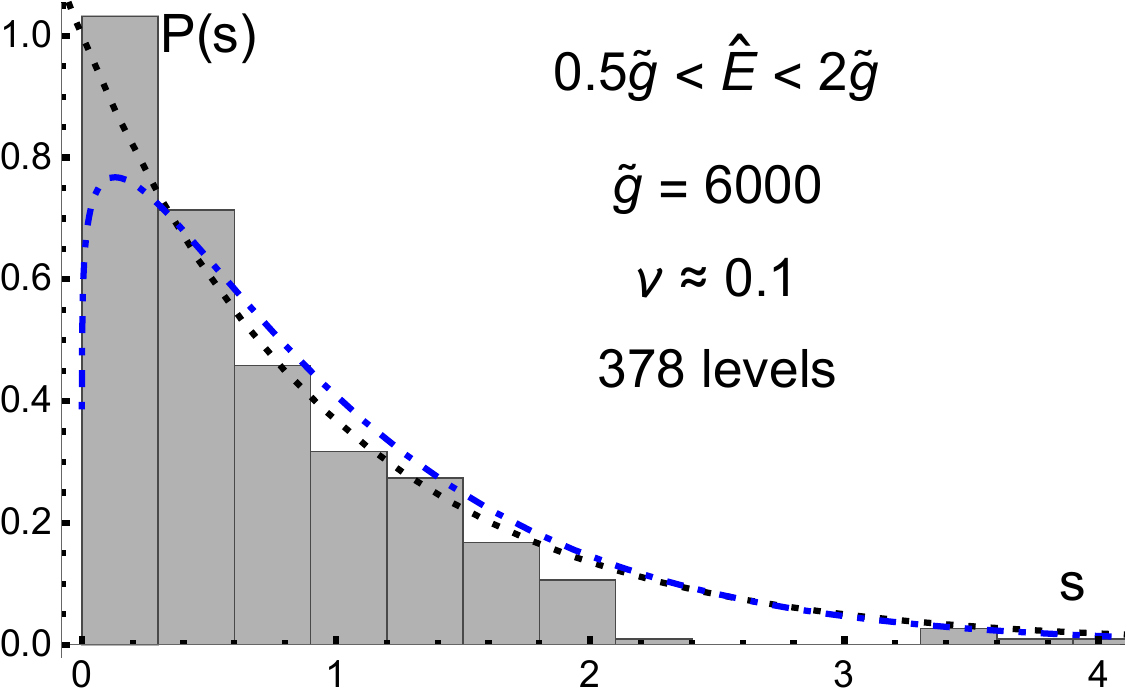}
	\caption{\small regular: exponential}
	\label{f:g-6k-e-pt5g-2g}
	\end{subfigure}
	\begin{subfigure}{0.49\columnwidth}
	\includegraphics[width = \textwidth]{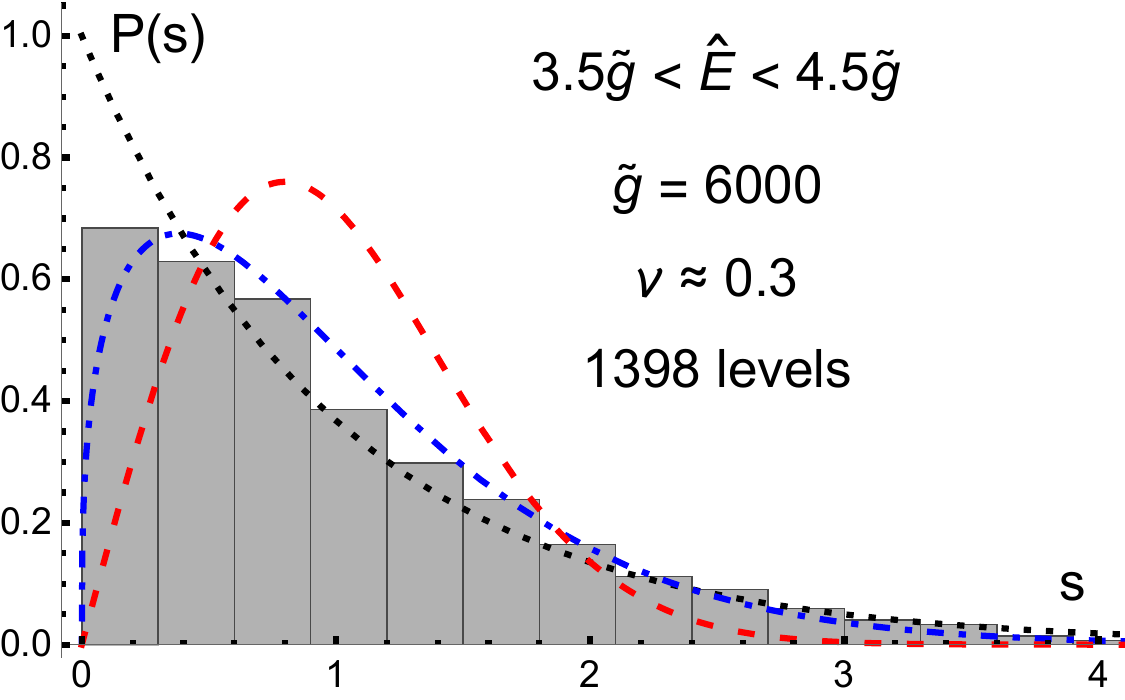}
	\caption{\small mixed}
	\label{f:g-6k-e-3pt5g-4pt5g}
	\end{subfigure}
	\begin{subfigure}{0.49\columnwidth}
	\includegraphics[width = \textwidth]{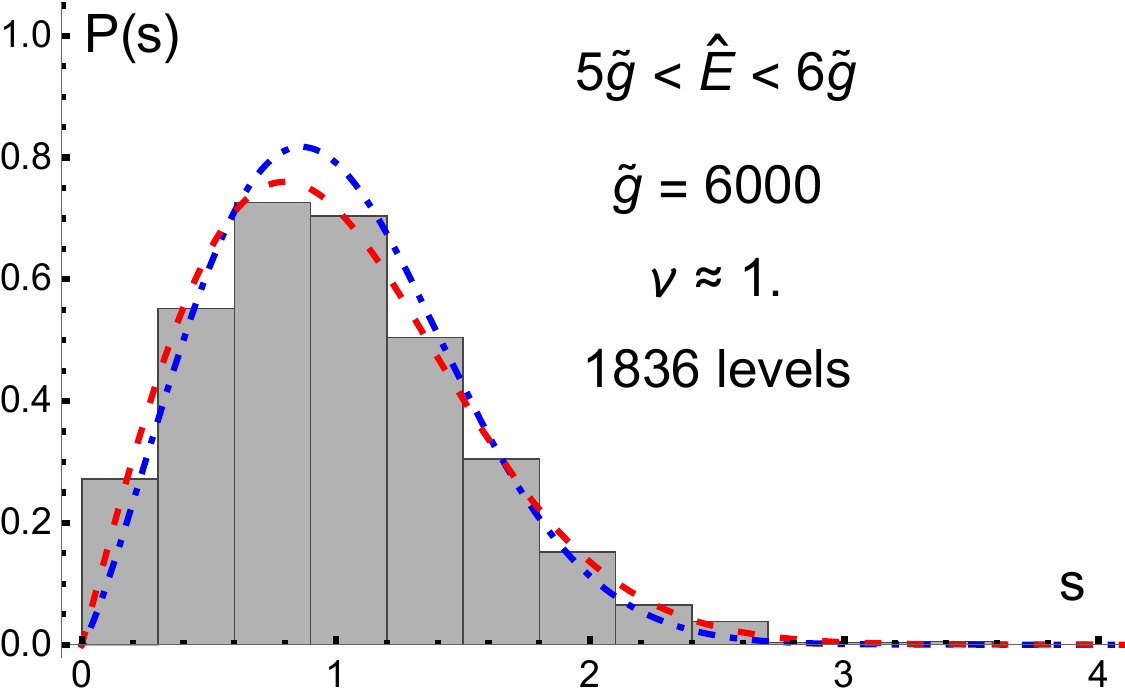}
	\caption{\small chaotic: Wigner}
	\label{f:g-6k-e-5g-6g}
	\end{subfigure}
	\begin{subfigure}{0.49\columnwidth}
	\includegraphics[width = \textwidth]{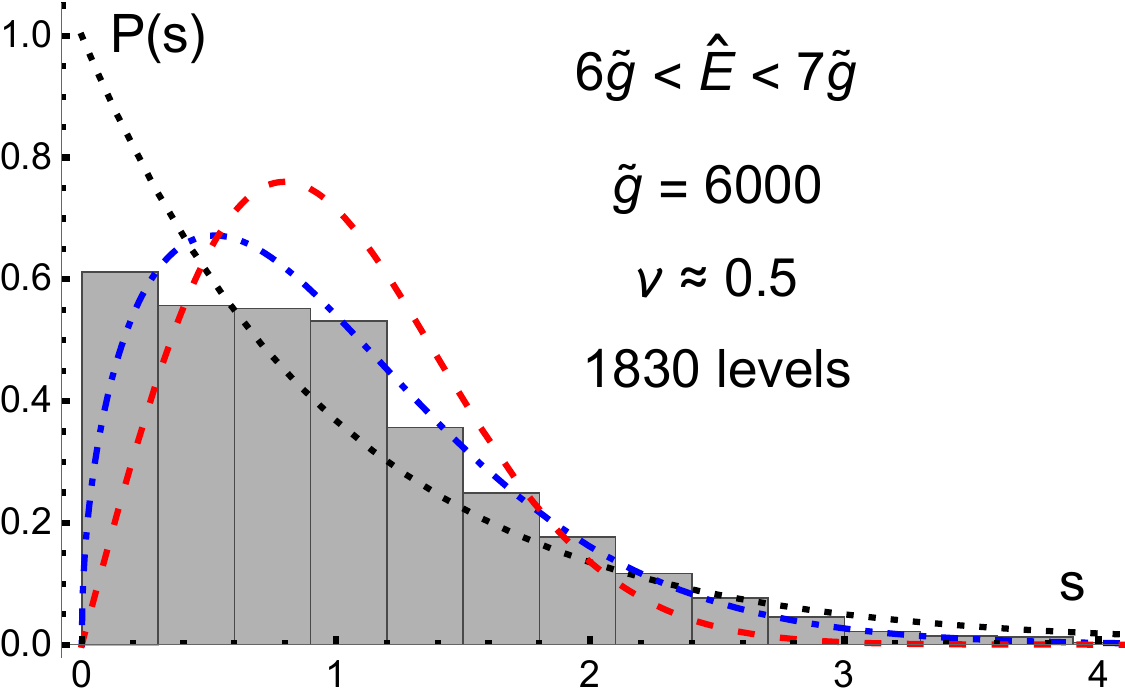}
	\caption{\small mixed}
	\label{f:g-6k-e-6g-7g}
	\end{subfigure}
	\begin{subfigure}{0.49\columnwidth}
	\includegraphics[width = \textwidth]{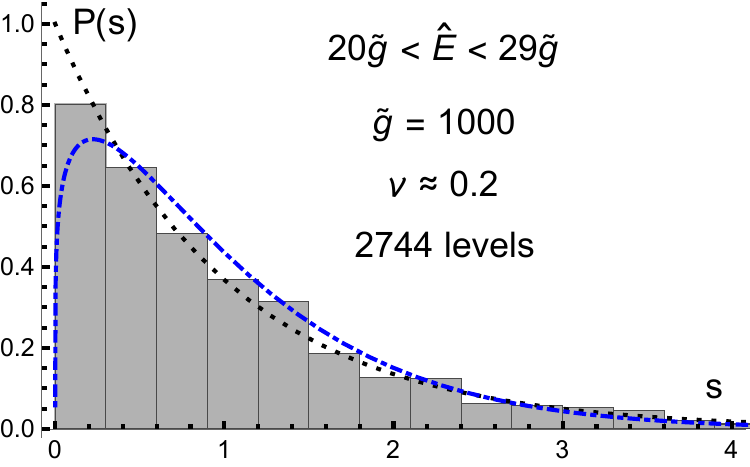}
	\caption{\small regular: exponential}
	\label{f:g-1k-e-20g-29g}
	\end{subfigure}
	\begin{subfigure}{0.49\columnwidth}
	\includegraphics[width = \textwidth]{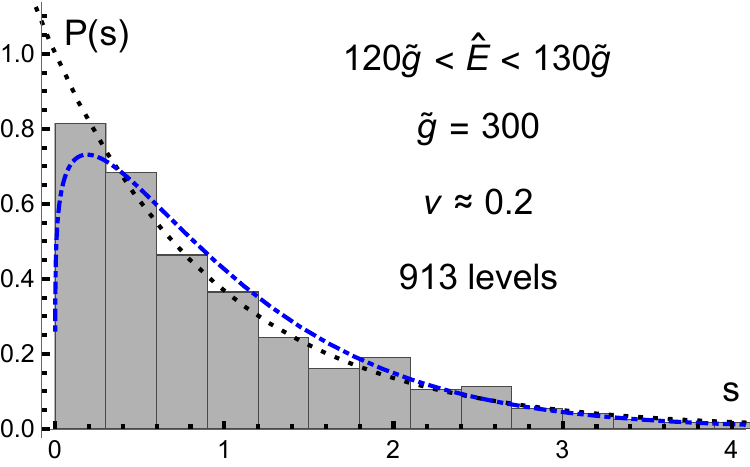}
	\caption{\small regular: exponential}
	\label{f:g-300-e-120g-130g}
	\end{subfigure}	
	\caption{\small Nearest neighbor spacing histograms (binwidth 0.1 for (a-b) and 0.3 for (c-g)) in various energy windows for $\tlg = 6000$ [in (g) and (h) we take $\tlg = 1000$ and $\tlg = 300$ due to inadequate high energy data for $\tlg = 6000$]. At low energies the spacing distribution in (b) is a perturbation to that of a desymmetrized/purified 2d harmonic oscillator (see Section \ref{s:harmonic-approx-purification-spacing}) shown in (a) which displays a large number of nearly degenerate levels. In (c-h) the spacing distributions are compared with exponential (black dotted), Wigner  surmise (red dashed) and Brody (blue dot-dashed) distributions showing an order-chaos-order transition. The fitted Brody parameter $\nu$ to one significant digit is indicated. Semiclassical expectations are seen to hold even when $\tlg$ is reduced to $300$.}
	\label{f:spacing-distrib}
\end{figure}

As shown in Fig.~\ref{f:spacing-distrib}, for $\tlg = 6000$, we observe (b) 2d harmonic oscillator-like spacing for $\hat E < 0.5 \tlg$ anticipated in Section \ref{s:harmonic-approx} and further discussed in Section \ref{s:harmonic-approx-purification-spacing}, (c) exponential spacing when $0.5 \tlg \leq \hat E \leq 2 \tlg$, (d) a Brody-like distribution with $\nu \approx 0.3$ for $3.5 \tlg \leq \hat E \leq 4.5 \tlg$, (e) Wigner-Dyson spacing for $5 \tlg \leq \hat E \leq 6 \tlg$ and (f) Brody with $\nu = 0.5$ for $6 \tlg \leq \hat E \leq 7 \tlg$. In (g) and (h) we see a return to regularity at higher energies ($20 \tlg \leq \hat E \leq 29 \tlg$ when $\tlg = 1000$ and $120 \tlg < \hat E < 130 \tl g$ when $\tlg = 300$). Thus, at least for large $\tlg$, we see a quantum analogue of the classical order-chaos-order behavior. Notably, while the classical degree of chaos (as quantified by the fraction of chaos estimated in Fig.~12 of \cite{gskhs-3rotor}) showed a rather sharp transition around $E \approx 3.8 g$, the quantum spacing distribution shows a more gradual transition for large $\tlg$. 

\subsection{Harmonic approximation: large \texorpdfstring{$\tlg$}{g-tilde} and low \texorpdfstring{$\hat E$}{E-hat}}
\label{s:harmonic-approx-purification-spacing} 

As mentioned in Section \ref{s:harmonic-approx}, for large $\tl g$, the states with low $\hat E$ may be captured by the harmonic approximation (\ref{e:harm-approx-hamil}). Prior to purification, the harmonic spectrum (\ref{e:harm-spectrum}) is 
	\beqs
	\frac{\hat E^{\rm harm}_{m,n}}{\sqrt{3 \tl g}} &=& m + n + 1 \quad \text{for} \quad m,n = 0,1,2, \ldots \cr 
	&=& 1, 2, 2, 3, 3, 3, 4, 4, 4, 4, 5, 5, 5, 5, 5, \cdots.
	\label{e:full-harmonic-spectrum}
	\eeqs
Evidently, the degeneracy is $m+n+1$. By comparing with the numerically obtained purified spectrum, we empirically deduce that the states in the identity representation have the energies
	\beq 
	\frac{\hat E^{\rm harm}_{\rm identity}}{\sqrt{3 \tl g}} = 1^1, 3^1, 5^1, 7^2, 9^2, 11^2, 13^3, 15^3, 17^3, 19^4, 21^4, \ldots
\label{e:purified-harmonic-spectrum-1st-few}
	\eeq 
with superscripts denoting degeneracies. In other words, from the full spectrum (\ref{e:full-harmonic-spectrum}), we select only levels with `odd' energies with the following frequency: $\hat E/\sqrt{3 \tl g} = 2 i + 1$ is chosen $\lceil \frac{2 i + 1}{6} \rceil$ times.  Here, $\lceil x \rceil$ is the smallest integer greater than or equal to $x$. The resulting spacing distribution (after unfolding) shown in Fig.~\ref{f:g-6k-e-0-pt5g-harmonic} accurately captures the spacings obtained from numerical diagonalization shown in Fig.~\ref{f:g-6k-e-0-pt5g}.

It would be useful to find a basis for the identity representation of $S_3 \times \mathbb{Z}_2$ in the harmonic approximation to check our proposed rule for purification. Prior to purification, the energy eigenfunctions [with energy $\hat E_{m,n} = \sqrt{3 \tlg} (m+n+1)$] in this approximation are
	\beqs
	\psi_{m,n}(\vf_+, \vf_-) &=&
	\psi^+_m(\vf_+) \psi^-_n(\vf_-) \quad	\text{where}\cr 
	\psi^+_m(\vf_+) &\propto& H_{m}((3 \tlg/4)^{1/4} \vf_+) \; e^{- \sqrt{3 \tlg/4} \vf_+^2 } \quad 	\text{and} \cr
	\psi^-_n(\vf_-) &\propto& H_{n}((\tlg/12)^{1/4} \vf_-) \; e^{- \sqrt{\tl g/12} \vf_-^2}.
	\eeqs
Here, $\vf_\pm$ are viewed as taking values on the real line although they arose as angles. This is a reasonable approximation since the Gaussian factors ensure that $\psi^+_m$ and $\psi^-_n$ decay rapidly. $H_n$ is the $n^{\rm th}$ Hermite polynomial. The exponent in the Gaussian factor in $\psi_{m,n}$ is invariant under the symmetry group $G = S_3 \times \mathbb{Z}_2$ since it is proportional to the potential $V(\vf_+,\vf_-)$ (\ref{e:harm-approx-hamil}). Hence, the desired basis consists of those linear combinations of $H_m H_n$ that are invariant under $G$. One may find them by group averaging: add to $H_m H_n$ all its translates under $G$. Since $G$ is a symmetry of the Hamiltonian, the orbit of $H_m H_n$ consists of states with the same energy. From (\ref{e:purified-harmonic-spectrum-1st-few}), we expect that 
\begin{enumerate}
\item[(i)] for even $\hat E/\sqrt{3 \tlg}$, this average vanishes so that there is no invariant combination, while 
\item[(ii)] for odd $\hat E/\sqrt{3 \tlg}$ there should be $\left\lceil \hat E/6\sqrt{3 \tlg} \right\rceil$ independent invariant energy eigenstates. 
\end{enumerate}

We now furnish a proof of (i). The action of $S_3 \times \mathbb{Z}_2$ on $\vf_\pm$ can be inferred from Table \ref{t:action-of-G-on-phipm}. Given the action of $\tau_{31} \in S_3$ and $\pi \in \mathbb{Z}_2$, the translates of $(\vf_+, \vf_-)$ may be organized as six pairs. The two translates in each pair differ only in the sign of the image of $\vf_+$: 
	\beqs
	&& (\pm \vf_+, \vf_-), \quad (\pm \vf_+, - \vf_-), \cr
	&& (\pm(\vf_+ - \vf_-)/2, (3\vf_+ +\vf_-)/2), \cr
	&& (\pm(\vf_+ - \vf_-)/2, - (3\vf_+ +\vf_-)/2), \cr
	&& (\pm(\vf_+ + \vf_-)/2, (3\vf_+ - \vf_-)/2), \cr
	&& (\pm(\vf_+ + \vf_-)/2, -(3\vf_+ - \vf_-)/2).
	\eeqs
Evidently, they may also be arranged so that each pair differ only in the sign of the image of $\vf_-$. Now we consider the group average of 
	\beq
	\psi_{m,n}(\vf_+,\vf_-) = \psi^+_m(\vf_+) \psi^-_n(\vf_-).
	\eeq	
Suppose $m$ is odd, so that $\psi^+_m$ is an odd function of $\vf_+$. Then the two terms from each pair in the above $G$-orbit cancel, leading to a vanishing group average. Analogously, the group average vanishes if $n$ is odd. Thus, there is no nontrivial $G$-invariant state in the $\hat E^{\rm harm}_{m,n}$ eigenspace if either $m$ or $n$ is odd. From (\ref{e:full-harmonic-spectrum}), for $\hat E_{m,n}/\sqrt{3 \tl g}$ to be even, either $m$ or $n$ must be odd. We thus conclude that in the identity representation, there cannot be any energy levels with even $\hat E/\sqrt{3 \tl g}$ within the harmonic approximation.

Although we have not yet found a general argument for (ii), we have verified our conjecture on degeneracies for low-lying levels with odd $\hat E/\sqrt{3 \tl g}$ in the identity representation. The first three levels $\hat E/\sqrt{3 \tl g} = 1,3,5$ are nondegenerate as we now show. The ground state has a constant $G$-invariant eigenfunction $\propto H_0 H_0 = 12$. The group averages of $H_2 H_0$ and $H_0 H_2$ are both equal to $4 (3 \sqrt{3} \vf_+^2 + \sqrt{3} \vf_-^2 - 6)$ while the group averages of $H_4 H_0, H_2 H_2$ and $H_0 H_4$ produce only one independent $G$-invariant state given by 6, 2 and 6 times
	\beq
	(24 - 8 \sqrt{3} \vf_-^2 + \vf_-^4 - 24 \sqrt{3} \vf_+^2 + 6 \vf_-^2 \vf_+^2 + 9 \vf_+^4).
	\eeq
For the fourth level with $\hat E = 7 \sqrt{3 \tl g}$, we find that the group averages of $H_0 H_6 \pm H_6 H_0$ are equal to those of $H_4 H_2 - H_2 H_4$ and $5(H_4 H_2 + H_2 H_4)$, leading to two linearly independent states, as expected from (\ref{e:purified-harmonic-spectrum-1st-few}). It would of course be satisfying to extend these special cases to a proof of the proposed degeneracy formula $\left\lceil \hat E/6\sqrt{3 \tlg} \right\rceil$ for odd $\hat E/\sqrt{3 \tl g}$.

\subsection{Spacings at low and moderate \texorpdfstring{$\tlg$}{g-tilde}}
\label{s:low-g-free-rotor}

\begin{figure}
\captionsetup[subfigure]{font=footnotesize,labelfont=footnotesize}
	\centering
	\begin{subfigure}{0.49\columnwidth}
	\includegraphics[width = \textwidth]{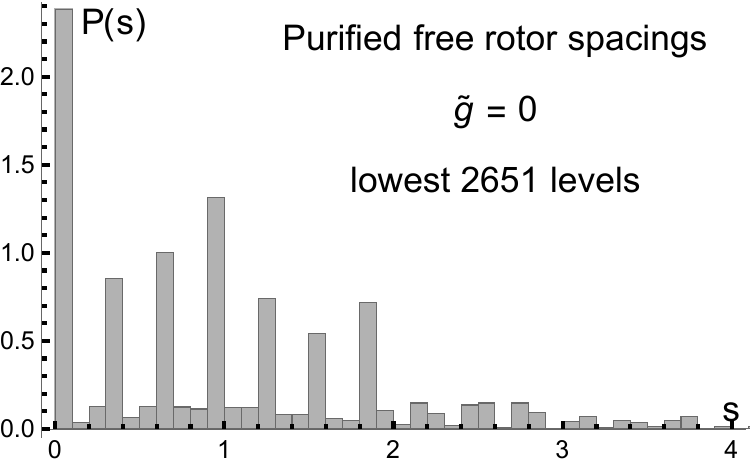}
	\caption{\small free rotors: $\tl g = 0$}
	\label{f:g-0-spacing}
	\end{subfigure}
	\begin{subfigure}{0.49\columnwidth}
	\includegraphics[width = \textwidth]{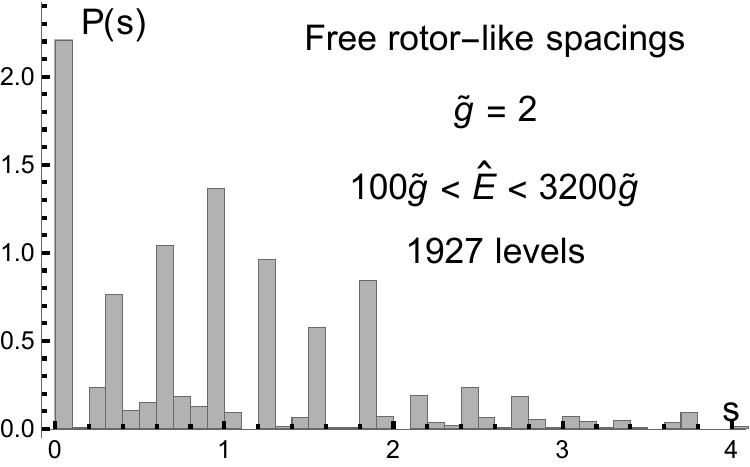}
	\caption{\small low $\tlg$, high $\hat E$}
	\label{f:g-2-e-100g-3200g}
	\end{subfigure}
	\begin{subfigure}{0.49\columnwidth}
	\includegraphics[width = \textwidth]{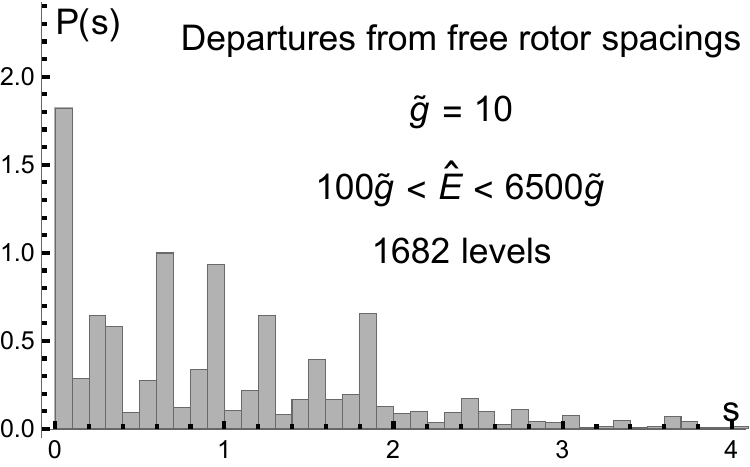}
	\caption{\small moderate $\tl g$, high $\hat E$ 
	}
	\label{f:g-10-e-100g-6500g}
	\end{subfigure}
	\begin{subfigure}{0.49\columnwidth}
	\includegraphics[width = \textwidth]{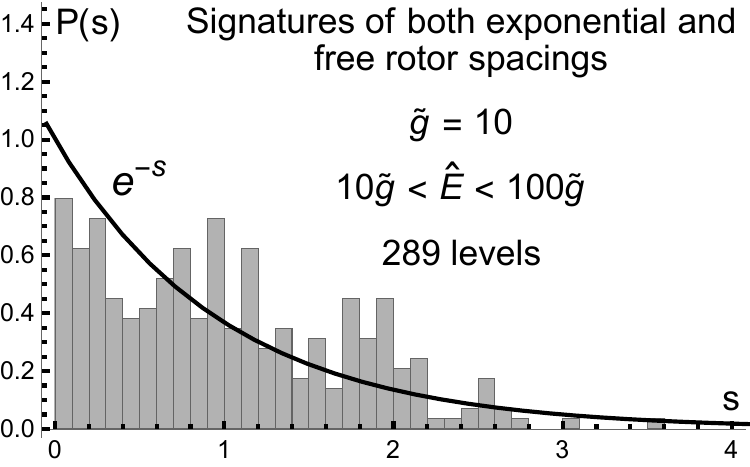}
	\caption{\small moderate $\tlg$ and $\hat E$ 
	}
	\label{f:g-10-e-10g-100g}
	\end{subfigure}
	\begin{subfigure}{0.49\columnwidth}
	\includegraphics[width = \textwidth]{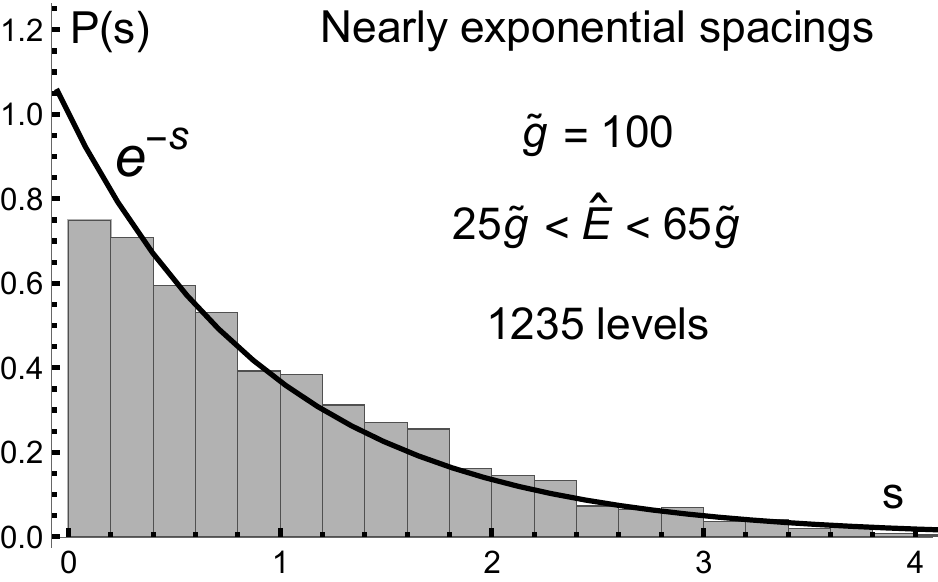}
	\caption{\small $\tlg = 100$, moderate $\hat E$}
	\label{f:g-100-e-25g-65g}
	\end{subfigure}
	\begin{subfigure}{0.49\columnwidth}
	\includegraphics[width = \textwidth]{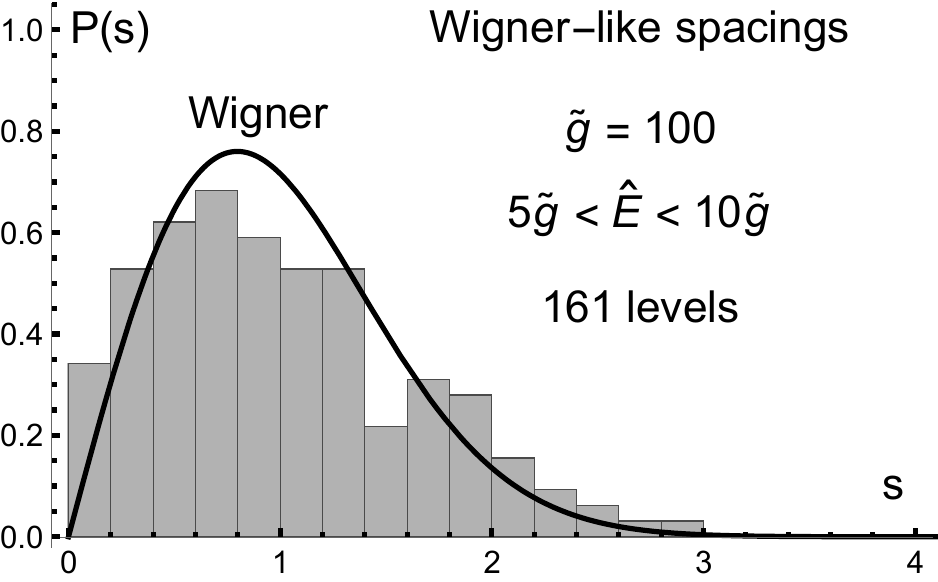}
	\caption{\small $\tlg = 100$, low $\hat E$}
	\label{f:g-100-e-5g-10g}
	\end{subfigure}
	\caption{\small Exact free rotor ($\tl g = 0$) spacing in the identity representation shown in (a) captures that of the highly excited levels for low $\tlg = 2$ shown in (b). For $\tlg = 10$, the highly excited levels in (c) show some departures from free rotor spacings while the moderately excited levels in (d) show features of both exponential and free rotor spacings. When $\tlg = 100$, spacings of moderately excited states in (e) is nearly exponential while those of intermediate levels in the classically chaotic regime is Wigner-like as shown in (f).} 
	\label{f:spacing-distrib-low-gtilde}
\end{figure}

When $\tl g$ is smaller, the main change is that there are very few energy levels in the range $0 \leq \hat E \leq 6 \tlg$ making it difficult to use spacing distributions to examine the quantum analog of the classical transition from regularity to chaos. For example, there are only 14 levels for $\tl g = 10$ and 105 levels for $\tl g = 100$ in this energy window. On the other hand, when $\hat E > 25 \tl g$, where the classical behavior is relatively regular, we may expect spacing distributions to display Poisson statistics. However, there can be departures from this expectation for at least two reasons: (i) the semiclassical approximation may break down for small $\tlg$ and (ii) when $\hat E \gg \tlg$, the dynamics should be a small perturbation to free rotor motion, for which the spacing distribution is different from exponential. In fact, the free rotor ($\tlg = 0$) energy eigenvalues in the identity representation are given by $\hat E_{m,n} = m^2 + n^2 - mn$ for $m,n$ satisfying (\ref{e:basis-label-range}) with eigenfunctions $s_{m,n}(\vf_1, \vf_2)$ leading to the spacing distribution shown in Fig.~\ref{f:g-0-spacing}. Pleasantly, for $\tl g = 2$, the spacings of highly excited unfolded levels shown in Fig.~\ref{f:g-2-e-100g-3200g} closely mimics the free rotor spacings. For $\tlg = 10$, departures from free rotor at high energies are more pronounced as seen in Fig.~\ref{f:g-10-e-100g-6500g}. By contrast, at lower energies in the classically regular regime, we begin to observe signatures of the exponential spacing distribution in addition to that of free rotors. This is visible in Fig.~\ref{f:g-10-e-10g-100g} for $\tlg = 10$. However, when $\tlg$ is increased to 100, the free rotor effects seem to become insignificant for these energies with spacing distributions approaching exponential as seen in Fig.~\ref{f:g-100-e-25g-65g}. Moreover, at $\tlg = 100$, we also begin to see Wigner-like spacing distributions in the classically chaotic regime (see Fig.~\ref{f:g-100-e-5g-10g}).

\section{Number variance and Spectral form factor}
\label{s:number-var-SFF}

\subsection{Number variance}
\label{s:number-var}

Let $d(\xi) = \sum_{i = 1}^N \del(\xi - \xi_i)$ denote the spectral density function for the unfolded energy levels $\xi_1 , \ldots, \xi_N$. Then the number of levels in a spectral window $[\xi - L/2, \xi+ L/2]$ is given by
	\beq
	n(\xi,L) = \int_{\xi-L/2}^{\xi + L/2} d(\xi') \: d \xi'.
	\eeq
The {\it number average} $\bra n(\xi,L) \ket$ is defined as the mean value of $n(\xi,L)$ as $\xi$ ranges over a desired portion of the spectrum. In practice, this ensemble average is performed by summing over an equally spaced grid of values of $\xi$ (with a spacing of $0.1$ for $L > 2$ and $0.05$ for $L \leq 2$). Since the unfolded spectrum is constructed to have approximately unit spacing on average, $\bra n(\xi, L) \ket \approx L$. Fluctuations in $n(\xi, L)$ are captured by the {\it number variance} \cite{mehta,stockmann} defined as
	\beq
	\Sig(L) = \bra [n(\xi,L) - \bra n(\xi,L) \ket]^2 \ket.
	\eeq
\begin{figure}
\captionsetup[subfigure]{font=footnotesize,labelfont=footnotesize}
	\centering
	\begin{subfigure}{0.48\columnwidth}
	\includegraphics[width = \textwidth]{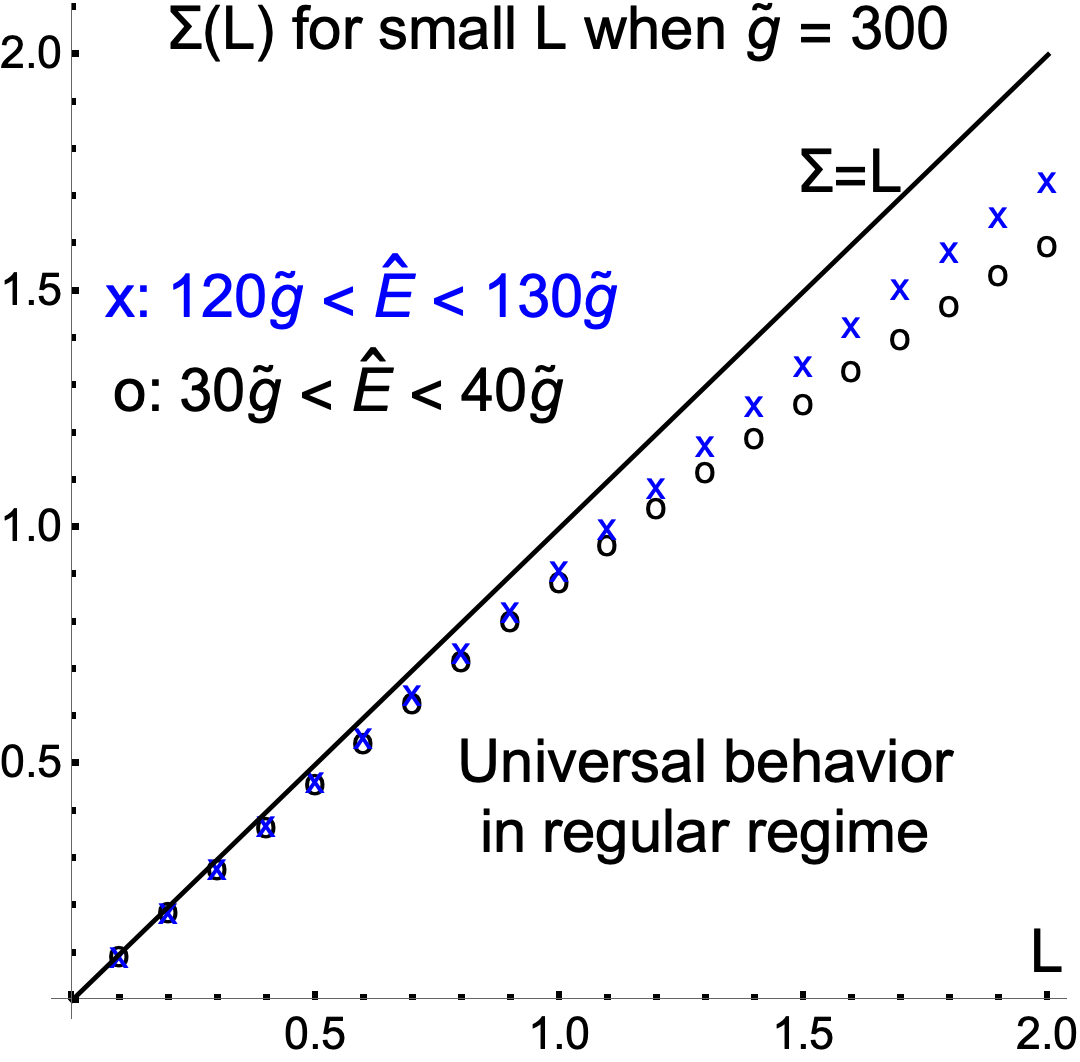}
	\caption{\small}
	\label{f:num-mean-integrable}
	\end{subfigure}
	\;
	\begin{subfigure}{0.48\columnwidth}
	\includegraphics[width = \textwidth]{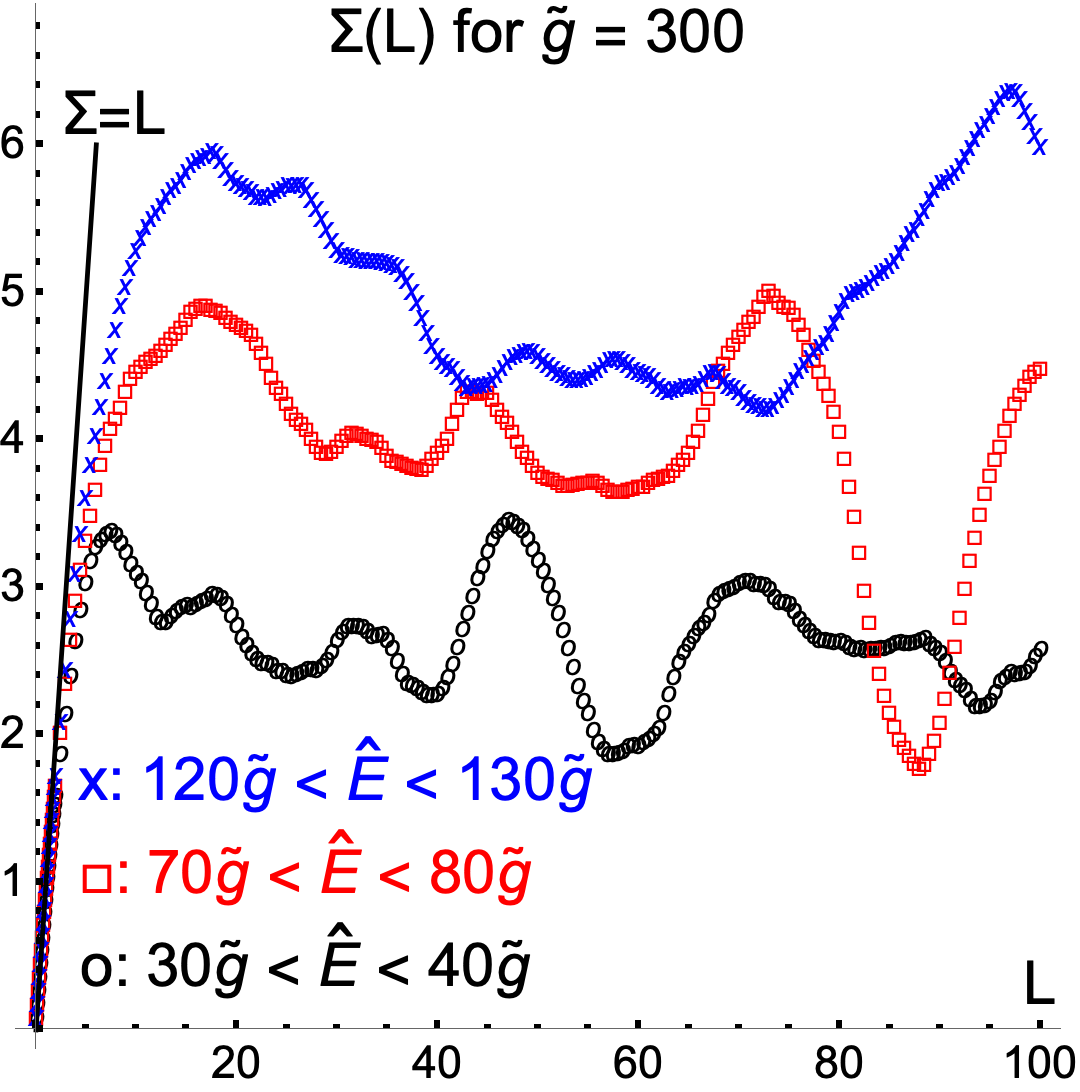}
	\caption{\small}
	\label{f:num-var-integrable}
	\end{subfigure}
	\caption{\small Number variance $\Sig(L)$ in classically regular regime of large $\hat E \gg \tlg$. (a) $\Sig(L)$ shows linear growth for small $L$ with a universal slope of unity. (b) For larger $L$, the number variance saturates and oscillates. The universal linear behavior persists up to higher values of $L$ at higher energies, where the dynamics approaches integrability.} 
	\label{f:number-var-reg-regime}
\end{figure}
\begin{figure}
\captionsetup[subfigure]{font=footnotesize,labelfont=footnotesize}
	\centering
	\begin{subfigure}{0.48\columnwidth}
	\includegraphics[width = \textwidth]{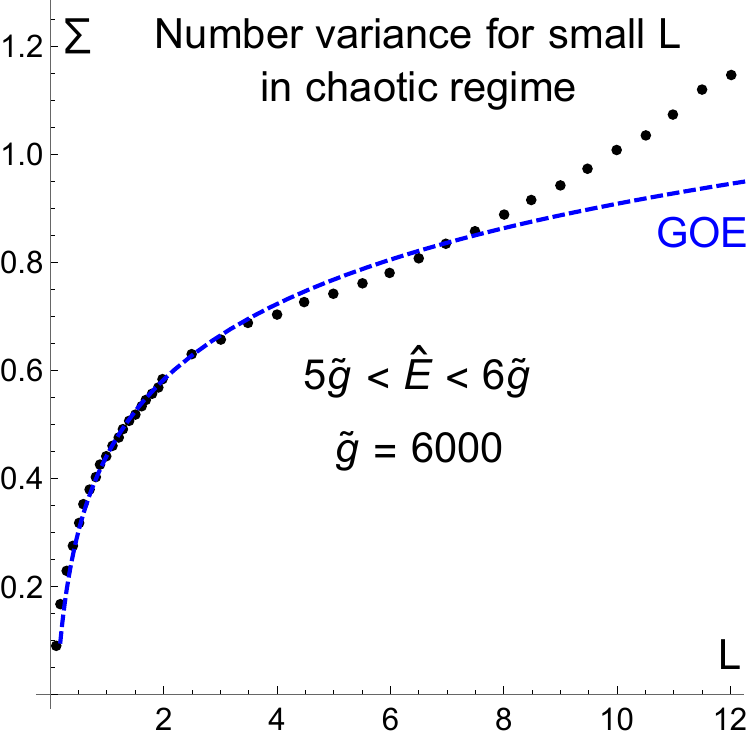}
	\caption{\small}
	\label{f:num-var-chaotic-g-6k-e-5g-6g-lowL}
	\end{subfigure} 
	\;
	\begin{subfigure}{0.48\columnwidth}
	\includegraphics[width = \textwidth]{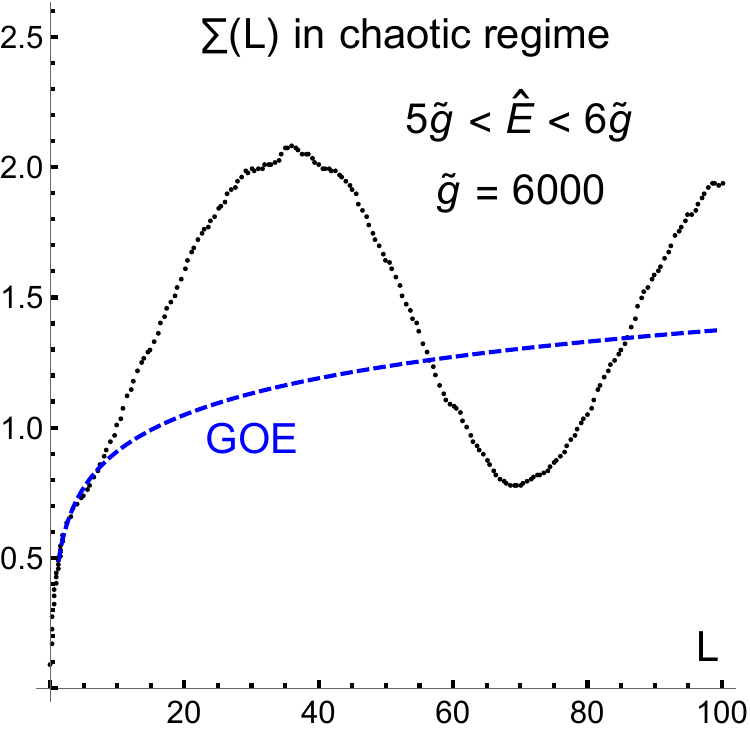}
	\caption{\small}
	\label{f:num-var-chaotic-g-6k-e-5g-6g}
	\end{subfigure}
	\caption{\small Number variance for 1836 levels in the classically chaotic energy band $(5 \tlg < \hat E < 6 \tlg)$ for $\tlg = 6000$ displays GOE asymptotics for moderately small $L$ ($0.1 \leq L \leq 4$) and then displays oscillations for larger $L$.} 
	\label{f:mean-and-number-variance-chaotic-regime}
\end{figure}
To interpret the number variance, we evaluate it for portions of the three-rotor spectrum in two representative regimes: (a) $1836$ levels with $5 \tlg < \hat E < 6 \tlg$ for $\tlg = 6000$ and (b) 921, 914 and 913 levels when $\hat E \in [30 \tlg, 40 \tlg$],  $[70 \tlg, 80 \tlg]$ and $[120 \tlg, 130 \tlg]$ for $\tlg = 300$ where level spacing distributions in Fig.~\ref{f:spacing-distrib} of Section \ref{s:semiclass-regime-large-g} indicate that the quantum dynamics are predominantly chaotic and regular respectively. It is conjectured that in these regimes, the number variance should show universal behavior for small $L$ that is captured by the Gaussian orthogonal ensemble (GOE) of random matrices and a Poisson process respectively. For the GOE, Dyson and Mehta (see (21) of \cite{dyson-mehta} and Appendix A38 of \cite{mehta}) derive the asymptotic behavior
	\beq
	\Sig(L) \sim \frac{2}{\pi^2} (\log(2\pi L) + \g + 1 - \pi^2/8) + {\cal O}(L^{-1}),
	\label{e:num-var-goe-dyson-mehta}
	\eeq
while for a Poisson process, $\Sig(L) = L$. Note that, though Mehta's logarithmic GOE formula is derived for asymptotically large $L$, while comparing with our number variance, it may be expected to hold in an intermediate regime: the ${\cal O}(L^{-1})$ remainder term dominates when $L \to 0$ while nonuniversal effects can become important for large $L$. We compare the number variance in the above regular and chaotic regimes against these expectations in Figs.~\ref{f:number-var-reg-regime} and \ref{f:mean-and-number-variance-chaotic-regime}. Although universal behavior is manifested for small/moderately small $L$, we find that in both the regimes, the number variance saturates and displays oscillations. Such oscillations are typical of several quantum systems \cite{casati-chirikov-guarneri-integ,aurich-steiner-num-var} and indeed even Riemann zeros \cite{berry-semicl-riemann-zero}, although the details are system specific. In fact, we would like to capture these oscillations in a semiclassical approximation by examining the short periodic orbits of the three-rotor system \cite{berry-spec-rigidity}.

\subsection{Spectral form factor}
\label{s:SFF}

\begin{figure}[]
\captionsetup[subfigure]{font=footnotesize,labelfont=footnotesize}
	\centering
	\begin{subfigure}{0.94\columnwidth}
	\includegraphics[width = \textwidth]{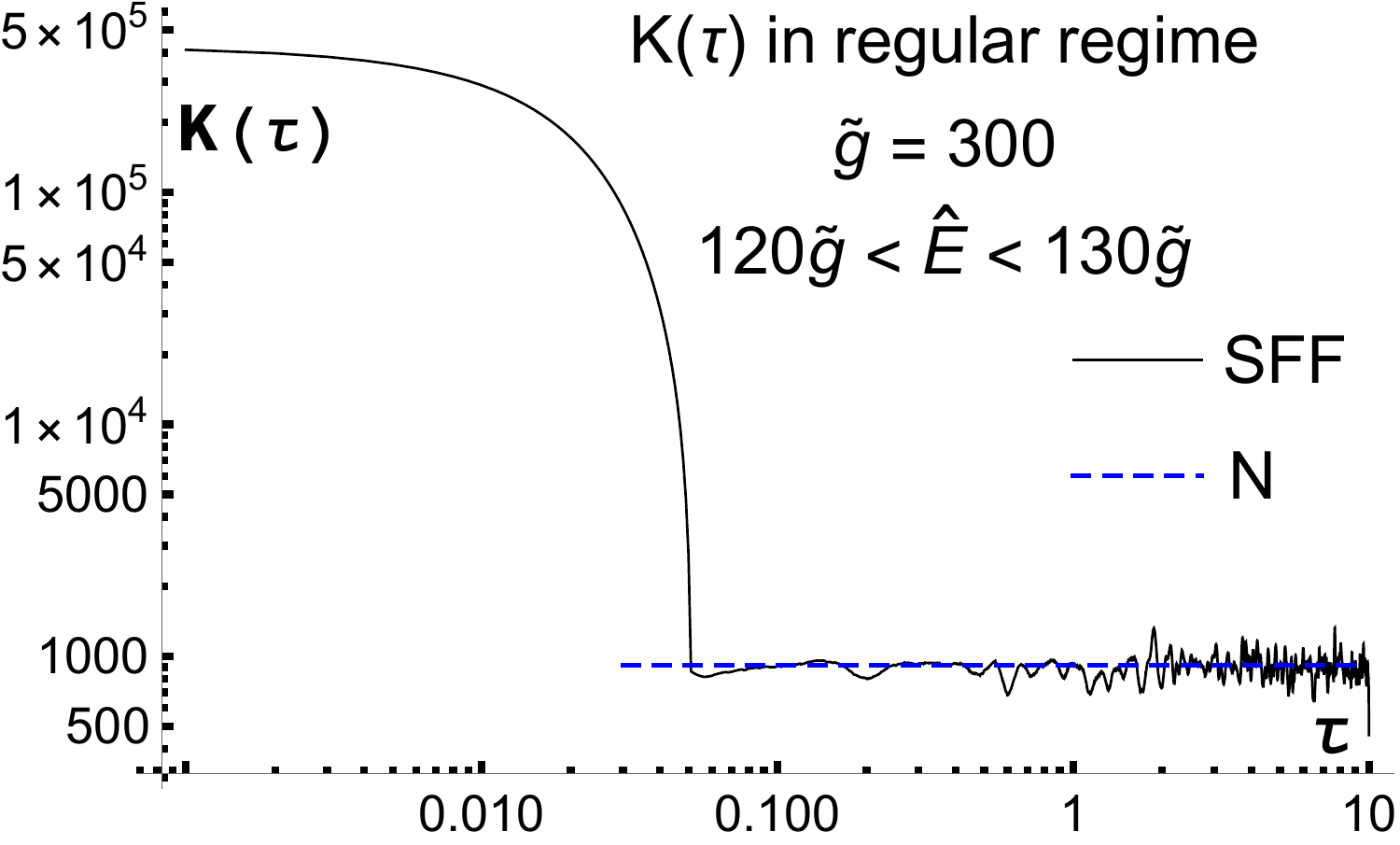}
	\caption{\small}
	\label{f:sff-e-120g-130g}
	\end{subfigure} 
	\;
	\begin{subfigure}{0.94\columnwidth}
	\includegraphics[width = \textwidth]{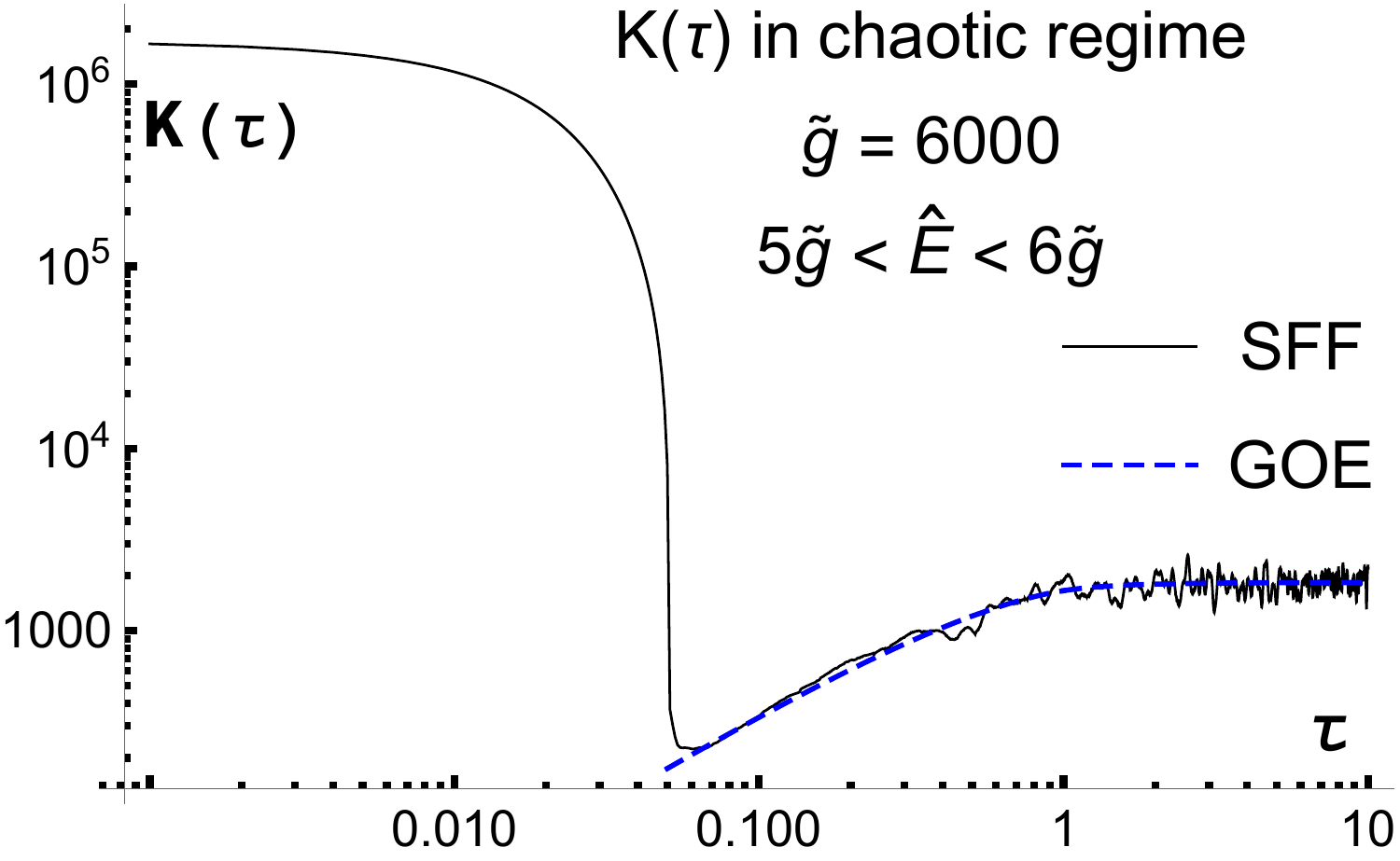}
	\caption{\small}
	\label{f:sff-e-5g-6g}
	\end{subfigure}
	\caption{\small Spectral form factor $K(\tau)$ on a log-log scale with limiting values $K(0) \sim N^2$ (reduced due to smoothing) and $K(\infty) = N$. (a) $N = 913$ unfolded levels in a classically regular energy range and (b) $N = 1836$ unfolded levels in a chaotic range. The SFF displays a nonuniversal broadened delta-like peak around $\tau = 0$. For moderate and large $\tau$, it displays nonuniversal high frequency fluctuations superimposed on a smooth profile equal to the number of levels $N$ in (a) and to the GOE prediction (\ref{e:goe-sff-formula}) in (b).} 
	\label{f:spec-form-fac-reg-chaotic}
\end{figure}

We now briefly report on the spectral form factor (SFF), which is another measure of two-point correlations in the energy spectrum \cite{haake,stockmann}. We define it as
	\beq
	K(\tau) = \bra Z^*(\tau) Z(\tau) \ket 
	\eeq
where $Z(\tau)$ is the Fourier sum
	\beq
	Z(\tau) = \sum_{n=1}^N e^{2 \pi i \xi_n \tau}. 
	\eeq
Here, $\xi_n$ are $N$ unfolded levels in a regular/chaotic/mixed energy window for a given value of $\tlg$. Moreover, $\tau = t/t_H$ where $t_H = 2\pi \hbar/\bra s \ket$ is the Heisenberg time. Since we have taken $\hbar = 1$ and the mean spacing of the unfolded spectrum is $\bra s \ket \approx 1$, $\tau = t/2\pi$. One sees that $Z(0) = N$ while due to cancellations, $Z(\tau)$ fluctuates around zero for nonzero $\tau$. On the other hand, $|Z(\tau)|^2$ displays a Dirac delta-like peak of height $N^2$ around $\tau = 0$ and is a rapidly fluctuating function of $\tau$ that tends asymptotically to $N$ as $\tau \to \infty$ \cite{levandier-lombardi-jost-pique-FT-stat-1986,chen-ludwig-sff-2018}. These high-frequency oscillations are nonuniversal and may be eliminated by averaging over several realizations or by applying a low-pass filter \cite{chen-ludwig-sff-2018,lozej-casati-prosen-sff-2022}. Since we have a single system, we perform the average $\bra \cdots \ket$ by convolving with a Gaussian smoothing kernel of width $0.05$. Although we do not do so here, we could simulate an ensemble of systems by considering several nearby values of $\tl g$. Some authors work with the connected two-point correlation ($K_c(\tau) = \bra |Z|^2 \ket - \bra Z \ket \bra Z \ket$) in which the delta-like peak at $\tau = 0$ should cancel out if the smoothing is performed carefully. We also note that (up to smoothing effects), the SFF is related to the number variance via an integral transform with respect to the square of the sinc function \cite{levandier-lombardi-jost-pique-FT-stat-1986}:
	\beq
	\Sigma(L) = \ov{N} \int_0^\infty d\tau K(\tau) \frac{2 \sin^2(\pi L \tau)}{(\pi \tau)^2}.
	\label{e:rel-betw-num-var-sff}
	\eeq


In Fig.~\ref{f:sff-e-120g-130g}, we plot the SFF in the classically regular regime $120 \tlg \leq \hat E \leq 130 \tl g$ for $\tlg = 300$. Aside from a nonuniversal broadened delta peak around $\tau = 0$, it displays rapid fluctuations (whose amplitude and details are system specific) about a constant value equal to the number of levels $N$ as expected \cite{stockmann} [see also (\ref{e:rel-betw-num-var-sff})].

In Fig.~\ref{f:sff-e-5g-6g}, we plot the SFF in the classically chaotic energy window $5\tlg \leq \hat E \leq 6 \tlg$ for $\tlg = 6000$. The graph of $K(\tau)$ displays (a) a broadened delta-like peak and dip around $\tau = 0$ followed by (b) a linear ramp-like increase and finally (c) a plateau-like saturation as $\tau \to \infty$. The delta peak of height $N^2$ is nonuniversal as are the high frequency fluctuations around the ramp and the plateau. On the other hand, the smooth parts of the ramp and plateau are captured by the GOE formula \cite{haake,lozej-casati-prosen-sff-2022}
	\beq
	K_{\rm GOE}(\tau) = N \times \begin{cases}
	2\tau - \tau \log(2\tau + 1) & \text{if} \;\; \tau < 1 \cr
	2 - \tau \log\left[ \frac{2\tau+1}{2\tau-1} \right]& \text{if} \;\; \tau > 1,
	\end{cases}
	\label{e:goe-sff-formula}
	\eeq
which is related to (\ref{e:num-var-goe-dyson-mehta}) via the integral transform (\ref{e:rel-betw-num-var-sff}). As with the number variance, it would be interesting to capture nonuniversal features of the SFF using properties of classical periodic orbits in a semiclassical approximation.




\section{Discussion}
\label{s:discussion}

Here, we discuss our results while highlighting some novel features of the quantum three-rotor problem and  mention some avenues for future research.

To begin with, the formulation of the quantum three-rotor problem involved some subtleties: singlevaluedness of the wave function allowed for nontrivial boundary conditions on the relative wave function (labelled by the cube roots of unity) with consequences for the center of mass angular momentum quantum number as well. The presence of a dimensionless coupling constant $\tl g$ distinguished the quantum theory from its classical counterpart. This was exploited in organizing the quantum theory into weak and strong coupling regimes with semiclassical behavior observed at strong coupling allowing us to interpret subsequent results on spectral statistics. However, even for large $\tl g$, making contact with universal semiclassical expectations in spectral statistics was complicated by the need to partition the spectrum into energy windows where the classical dynamics is regular, mixed or chaotic. By contrast, in models such as the planar elastic pendulum \cite{sagar-chakraborty-order-chaos-order-PEP, sagar-chakraborty-quantum-PEP} which also displays order-chaos-order behavior, there is a control parameter (other than energy) whose variation shifts the entire spectrum from Poisson to Wigner-Dyson.

Another interesting feature of the three-rotor model is the presence of an $S_3 \times \mathbb{Z}_2$ discrete symmetry of the relative Hamiltonian. Although this symmetry is manifest in the rotor angle description, it is not obvious while working with the relative or Jacobi angles. A key step was to identify the implementation of this symmetry on relative angles in order to be able to decompose the energy spectrum into irreducible representations where universal behavior may be expected. In fact, incorrect desymmetrization/purification via a more obvious $\mathbb{Z}_2 \times \mathbb{Z}_2$ symmetry would lead to spurious Shnirelman-like peaks \cite{shnirelman-1975, chirikov-shepelyansky-1995} in spacing distributions. 

Group averaging allowed us to construct a convenient basis for the subspace of the Hilbert space that carries the identity representation of $S_3 \times \mathbb{Z}_2$, allowing us to purify the spectrum. Pleasantly, we also have a closed-form sparse matrix representation of the Hamiltonian in this basis. Moreover, we demonstrate that numerically determined eigenvalues are insensitive to an increase in the size of the truncated Hamiltonian ensuring that spacing distributions are computed to desired accuracy. Although we have found a basis  for the identity representation, we suspect that there is a more elegant way of choosing orbit representatives to label the basis elements. In fact, the problem of finding such a basis for this as well as for the remaining five irreducible representations is an interesting mathematical problem related to Schur polynomials and generalizations studied by representation theorists \cite{macdonald-schur}. As the other representations of our symmetry group $S_3 \times \mathbb{Z}_2$ are real, we would expect the corresponding universal spectral statistics to follow GOE predictions \cite{keating-gue-goe}. Nevertheless, it would still be interesting to find bases for these representations and study departures from universal behavior.

Remarkably, the above $S_3 \times \mathbb{Z}_2$ symmetry continues to apply even at the level of the harmonic approximation, which allowed us to use the latter to predict the spacing distributions at low energies where there are marked departures from the semiclassical Poisson distribution. However, while we have conjectured which harmonic levels belong to the identity representation (along with their degeneracies), we have not yet found general expressions for suitable basis elements for the invariant subspace in the harmonic approximation (unlike in the full problem where we have explicit formulae for an orthonormal basis). 

On the other hand, we showed that the free rotor approximation captures  the (non-Poisson) spacing statistics at asymptotically high energies and weak coupling. Interestingly, finding the degeneracies of the free rotor spectrum led us to a number theoretic problem involving the Eisenstein integers, whose solution could be expressed in terms of the prime factorization of the energy in suitable units. 

There are several other interesting directions for further research such as to understand (i) the nature of level crossings and avoided crossings, (ii) higher order spacing ratios, especially as a tool to assist in the decomposition of the spectrum \cite{tekur-bhosale-santhanam} into irreducible representations, (iii) the behavior of wave functions, Husimi functions and their nodal curves in the regular and chaotic regimes, (iv) the saturation and oscillation of the number variance $\Sig(L)$ for large $L$ and the nonuniversal behavior of the spectral form factor using our knowledge of classical periodic orbits, and (v) out-of-time-ordered correlators (OTOC) and their connection to classical Lyapunov exponents.


\begin{acknowledgments} 

We thank M V Berry, S R Jain, A Lakshminarayan, M S Santhanam and anonymous reviewers for corrections, helpful comments and references. We thank G D Patil for guiding us through the number theory behind the degeneracy formula for free rotor energy levels given in Section \ref{s:pert-theory} and Appendix \ref{a:degeneracy-formula-unperturbed}. Finally, although Ankit Yadav did not wish to be included as an author, we are grateful to him for his valuable contributions at an intermediate stage of this work. This work was supported in part by the Infosys Foundation and grants (MTR/2018/000734, CRG/2018/002040) from the Science and Engineering Research Board, Govt. of India.

\end{acknowledgments}

\appendix

\section{Free rotor degeneracy and Eisenstein primes}
\label{a:degeneracy-formula-unperturbed}

Recall from Section \ref{s:pert-theory} that the spectrum of the free rotor Hamiltonian $\hat H = T$ is given by $m^2 + n^2 - mn$ where $m$ and $n$ are integers. Here, we outline a proof of formula (\ref{e:degeneracy-free-rotor-levels}) for the degeneracy of free rotor energy levels based on standard number theoretic techniques such as those discussed in \cite{cox-primes}: if the prime factorization of the energy eigenvalue is of the form $T = 3^\g p_1^{\al_1} \cdots p_r^{\al_r} q_1^{\beta_1} \cdots q_s^{\beta_s}$, where $p_i \equiv 1$ $(\mod 3)$ and $q_i \equiv 2$ $(\mod 3)$, then the number of integer solutions to $m^2 + n^2 - mn = T$ is 
	\beq
	\text{degen}(T) = 6 (\al_1 + 1) (\al_2 + 1) \cdots (\al_r + 1)
	\label{e:degeneracy-T-appendix}
	\eeq
if all $\beta_i$ are even and $\text{degen}(T) = 0$ otherwise. After this paper appeared as a preprint, S R Jain informed us that a similar result was presented by Itzykson and Luck while studying arithmetic degeneracies in triangular billiards \cite{itzykson-luck}. However, they restrict to $m > n > 0$, so that the prefactor $6$ is absent. Moreover, their approach is somewhat analytic, being based on generating functions while ours is more algebraic.

To begin with, the factorization 
	\beqs
	m^2 + n^2 - mn &=& |m + n \om|^2 = (m + n \om)(m + n \om^2) \cr
	&=& (m + n \om) ((m-n) - m \om),
	\label{e:factorize-T}
	\eeqs
where $\om$ is a primitive cube root of unity suggests that the integral lattice spanned by $1$ and $\om$ is relevant to this problem. Thus, let $\om$ denote a complex number satisfying $\om^2 + \om + 1 = 0$ and let $L = \mathbb{Z}[\om] = \mathbb{Z} + \mathbb{Z} \om = \{ \: m + n \om | m,n \in \mathbb{Z} \}$ be the triangular lattice in the complex plane displayed in Fig.~\ref{f:triangular-lattice}. The set $L$ is closed under addition and multiplication and forms a ring called the Eisenstein integers, the ring of integers in $\mathbb{Q}(\om)$, which is an imaginary quadratic (same as $\mathbb{Q} + \mathbb{Q}(\sqrt{-3})$) and cyclotomic number field \cite{cox-primes}. The labels in Fig.~\ref{f:triangular-lattice} correspond to the choice $\om = e^{2\pi i/3}$, although one gets the same triangular lattice upon choosing $\om = e^{4\pi i/3}$. 

Let us start with some terminology and basic facts about $L$. {\bf (i)} Given $r, s \in L$, we say that $s$ divides $r$ (denoted $s|r$) if there is some $q \in L$ such that $r = s q$. {\bf (ii)} The elements of the set 
	\beqs
	U &=& \{ 1, -\om^2 = \om + 1, \om, -1, \om^2 = - \om - 1, -\om \} \cr
	&=& \{ (-\om)^k, k = 0,1, \ldots, 5 \}
	\eeqs
lying at the vertices of the hexagon marked in Fig.~\ref{f:triangular-lattice} of the lattice constitute the group of units of $L$. They behave like $\pm 1$ in the ring of integers $\mathbb{Z}$. More precisely, if $xy = 1$ for $x,y \in L$ then $x,y \in U$. Furthermore, units are the only elements of $L$ of unit magnitude: $|x| = 1$ implies $x \in U$. {\bf (iii)} An element $q$ that is not a unit ($q \in L \setminus U$) is called irreducible if in any factorization $q = ab$, at least one of the factors is a unit ($a \in U$ or $b \in U$). On the other hand, $q \in L \setminus U$ is called a prime if whenever $q | rs$ either $q | r$ or $q | s$. While primes are always irreducible, one may show that all irreducible elements of $L$ must be primes as $L$ is a Unique Factorization Domain (see below).

Using these definitions, we state the following division algorithm. Given a dividend $a \in L$ and a nonzero divisor $b \in L$, there are uniquely determined quotient and remainder elements $q, r \in L$ with $a = b q + r$ where $0 \leq |r| \leq |b/2|$ or $r = b/2$. In other words, $L$ is a norm Euclidean ring. Consequently, $L$ is a Principal Ideal Domain and hence a Unique Factorization Domain (Corollary 4.4 of \cite{cox-primes}). Moreover, from Proposition 4.7 of \cite{cox-primes}, the primes in $L$ (called Eisenstein primes) must be of one of the following sorts up to units: (i) $1 + 2 \om$ or (ii) $m + n \om$ or $m + n \om^2$ where $m^2 + n^2 - mn$ is an ordinary prime $\equiv 1$ $(\mod 3)$ in $\mathbb{Z}$ or (iii) an ordinary prime $q \equiv 2$ $(\mod 3)$ in $\mathbb{Z}$.

We now return to the integers and explain the degeneracy formula (\ref{e:degeneracy-T-appendix}). For given kinetic energy eigenvalue $T$, suppose $(m,n)$ is one integer pair such that $T = m^2 + n^2 - mn = |m + n \om|^2$. Then since the six units $u \in U$ have unit magnitude, $T = |u(m + n \om)|^2$ where $u(m + n \om) = m' + n' \om \in L$. Thus we get five additional distinct integer pairs with the same eigenvalue $T$. This explains the prefactor $6$ in (\ref{e:degeneracy-T-appendix}). 

It remains to find the contribution to the degeneracy that does not arise from the freedom to multiply by units. To do this, we will exploit the prime factorization (in $\mathbb{Z}$) $T = 3^\g p_1^{\al_1} \cdots p_r^{\al_r} q_1^{\beta_1} \cdots q_s^{\beta_s}$, where $p_i \equiv 1$ $(\mod 3)$ and $q_i \equiv 2$ $(\mod 3)$. The general idea is that any representation of $T$ as $|m + n \om|^2 = (m + n \om)(m + n \om^2)$ must arise from a product of such representations of the individual prime factors. The paired appearance of $m + n \om$ and $m + n \om^2$ is because $\om^2$ is the complex conjugate of $\om$. This is analogous to the nonreal roots of a polynomial with real coefficients appearing in complex conjugate pairs. We will now deal sequentially with the three sorts of primes in $\mathbb{Z}$: $3$, $p \equiv 1$ $(\mod 3)$ and $q \equiv 2$ $(\mod 3)$.

{\bf (a)}  We begin with the ordinary prime $3$ and note that $3 = |1 + 2 \om|^2 = (1 + 2 \om)(1 + 2 \om^2) = -(1 + 2 \om)^2$. Thus although $3$ is a prime in $\mathbb{Z}$, it is not a prime in $L$. In fact, $3$ ramifies in $L$ with ramification index two since it is a unit times the square of the prime $1 + 2 \om$ in $L$. Moreover, up to multiplication by units, there is only one way of expressing $3$ as $|m + n \om|^2$. Thus, even if the prime factorization of $T$ includes $3^\g$ for some $\g \geq 1$, there is only one way to write $3^\g$ as $(m+n\om) \times (m+n\om^2)$: $(1+2\om)^\g \times (1+2\om^2)^\g$. Hence, it will only contribute a factor of one to $\text{degen}(T)$. 

{\bf (b)} Suppose $p \equiv 1$ $(\mod 3)$ is a prime in $\mathbb{Z}$, then there exist integers $m, n$ such that $p = m^2 + n^2 - mn$. For example, $p = 7, 13$ and $19$ are primes $\equiv 1$ $(\mod 3)$ with $7 = |1 + 3 \om|^2 = |3 + \om|^2$, $13 = |1 + 4 \om|^2 = |4 + \om|^2$ and $19 = |2 + 5 \om|^2 =  |5 + 2 \om|^2$. An inductive argument can be used to show that the same applies to any other such prime. Moreover, in each case, there are precisely two representations $p = |m + n \om|^2 = |\tl m + \tl n \om|^2$ that do not differ by a unit but are related by $\tl m + \tl n \om = (-\om)^k (n + m \om)$ for some $0 \leq k \leq 5$. This is a reflection of the Galois symmetry $\om \to \om^2$ that matches the complex conjugation $i \to -i$. More generally, if $T = p^\al$ where $\al \geq 1$ and $p \equiv 1$ $(\mod 3)$, then (aside from the freedom to multiply by units) there are $\al + 1$ distinct ways of writing $T = |m + n \om|^2$. To see this, we denote $A = m + n \om$ and $B = n + m \om$. Then $p^2 = |A^2|^2 = |AB|^2 = |B^2|^2$ and in general, $p^\al = |A^\al|^2 = |A^{\al-1} B|^2 = \cdots = |A B^{\al - 1}|^2 = |B^\al|^2$. 


{\bf (c)} Finally, suppose $q$ is a prime number in $\mathbb{Z}$ and $q \equiv 2$ $(\mod 3)$. Then $q$ is also an Eisenstein prime. Thus, if $T = q^\beta$, then: (i) if $\beta$ is odd, there is no way of writing $T$ as $|m + n \om|^2$ but (ii) if $\beta$ is even, there is only way of writing $T$ as $|m + n \om|^2$ i.e., by taking $m + n \om = q^{\beta/2}$ up to a unit. This generalizes to $T = q_1^{\beta_1} \cdots q_s^{\beta_s}$. 

Combining these, we arrive at the degeneracy formula (\ref{e:degeneracy-T-appendix}).

\section{Basis labels for identity representation}
\label{a:orth-basis}

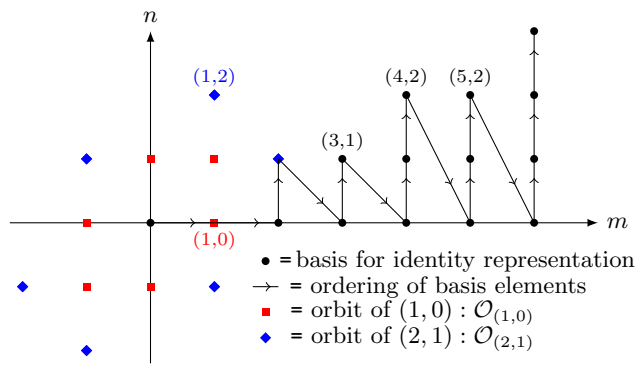
\begin{figure}
\centering
\begin{tikzpicture}[scale = .85]
\coordinate (Origin)   at (0,0);
\coordinate (XAxisMin) at (-2.2,0);
\coordinate (XAxisMax) at (7,0);
\coordinate (YAxisMin) at (0,-2.2);
\coordinate (YAxisMax) at (0,3);
\draw [thin,-latex] (XAxisMin) -- (XAxisMax) node[right] {$m$};
\draw [thin,-latex] (YAxisMin) -- (YAxisMax) node[above] {$n$} ;

\def\m{1};
\def\n{1};
\foreach \p/\q in {\m/\n,\n/\m,\n-\m/\n,\m/\m-\n,-\n/-\m,-\n/\m-\n,\n-\m/-\m,-\m/-\n,\m-\n/-\n,-\m/\n-\m,\n/\n-\m,\m-\n/\m}{\node[fill,rectangle,red,inner sep=0pt,minimum size=3pt] at (\p,\q) {};}

\def\m{2};
\def\n{1};
\foreach \p/\q in {\m/\n,\n/\m,\n-\m/\n,\m/\m-\n,-\n/-\m,-\n/\m-\n,\n-\m/-\m,-\m/-\n,\m-\n/-\n,-\m/\n-\m,\n/\n-\m,\m-\n/\m}{\node[fill,rectangle,rotate=45,blue,inner sep=0pt,minimum size=3pt] at (\p,\q) {};}

\node[above] at (3,1){\scriptsize (3,1)};
\node[above] at (4,2){\scriptsize (4,2)};
\node[above] at (1,2){\scriptsize \Blue (1,2)};
\node[below] at (1,0){\scriptsize \Red (1,0)};
\node[above] at (5,2){\scriptsize (5,2)};

\node[fill, circle, black, inner sep=0pt,minimum size=3pt] at (1.8,-.6){};
\node[] at (4.8,-.6){\small $\text{\texttt{=}\,basis for identity representation}$};

\node[] at (4.5,-1){\small $\text{\texttt{=} ordering of basis elements}$};

\node[fill, rectangle, red, inner sep=0pt,minimum size=3pt] at (1.8,-1.4){};
\node[] at (4.1,-1.4){\small $\text{\texttt{=}  orbit of } (1,0): \mathcal{O}_{(1,0)}$};

\node[fill, rectangle, blue, rotate=45, inner sep=0pt,minimum size=3pt] at (1.8,-1.8){};
\node[] at (4.1,-1.8){\small $\text{\texttt{=} orbit of } (2,1): \mathcal{O}_{(2,1)}$};

\begin{scope}[decoration={
    markings,
    mark=at position 0.7 with {\arrow{>}}}]
\node[fill,circle,inner sep=0pt,minimum size=3pt] at (0,0){};
\node[fill,circle,inner sep=0pt,minimum size=3pt] at (2,0){};
\draw[postaction={decorate}] (1.6,-1)--(2,-1);
\draw[postaction={decorate}] (0,0)--(1,0);
\draw[postaction={decorate}] (1,0)--(2,0);
\draw[postaction={decorate}] (2,0)--(2,1);
\foreach \X/\Y[remember=\X as \lastX (initially 2), remember=\Y as \lastY (initially 1)] in {3/0,3/1,4/0,4/1,4/2,5/0,5/1,5/2,6/0,6/1,6/2,6/3}
 {\draw[postaction={decorate}] (\lastX,\lastY)--(\X,\Y);
\node[fill,circle,inner sep=0pt,minimum size=3pt] at (\X,\Y) {};}
\end{scope}

\end{tikzpicture}
	\caption{\small Basis labels for identity representation of $S_3 \times \mathbb{Z}_2$ are orbit representatives $(m,n)$ (denoted by black dots) lying in a wedge in the $\mathbb{Z} \oplus \mathbb{Z}$ lattice. Each orbit intersects the wedge exactly once.}
	\label{f:wedge-basis-identity-representation}
\end{figure}

As mentioned in Table \ref{t:action-of-S3-Z2-on-angles}, the symmetry group $G$ acts on the Fourier basis states $e_{m,n} $ which may be viewed as points on the lattice of ordered pairs of integers $(m,n)$. The orbit of $(m,n)$ is \small
	\beqs
	{\cal O}_{(m,n)} &=& \{( m, n), (n - m, n), (m, m-n), (-n, -m),\cr 
	&& (-n, m-n), (n-m, -m), (-m, -n), (m-n, -n),\cr
	&& (-m, n-m), (n, m), (n, n-m), (m-n, m) \}.
	\label{orb-mn}
	\eeqs \normalsize
It is typically of length twelve except in the extreme cases [boundary of the wedge defined in (\ref{e:lat-cond}) and displayed in Fig.~\ref{f:wedge-basis-identity-representation}] when (a) $m = n=0$ and it shrinks to a single point and when (b) ($m>0$, $n= 0$ or $\lfloor m/2 \rfloor$) and it has length six. Now, we show that each orbit can be labelled uniquely by a representative $(m,n)$ lying in the `wedge' defined by
	\beq
	m \ge 0 \quad \text{and} \quad 0 \le n \le \lfloor m/2 \rfloor,
	\label{e:lat-cond}
	\eeq
where $\lfloor \cdot \rfloor$ is the greatest integer part. The proof proceeds in two steps: (i) We use the group action to argue that every orbit enters the wedge and (ii) show that an orbit cannot have more than one representative in the wedge.

\fl {\bf Proof of (i).} Of the points $(m,n)$ and $(-m,-n)$ lying in ${\cal O}_{(m,n)}$ [see (\ref{orb-mn})] one must lie in the right half plane. Without loss of generality we suppose that $(m,n)$ lies in the right half plane, i.e., $m \ge 0$. Next, one of the two orbit elements $(m,n)$ or $(m-n,-n)$ must lie in the upper half plane. Taking $(m,n)$ to lie in the upper half plane, we get $n \ge 0$ allowing us to restrict $(m,n)$ to the first quadrant. In a similar manner, either $(m,n)$ or $(n,m)$ must lie in the south-east half plane allowing us to restrict to the $\pi/4$ wedge: $m \ge n \ge 0$. Finally, either $(m, m-n)$ or $(m,n)$ must lie in the $\pi/8$ wedge of Eqn. (\ref{e:lat-cond}). Thus we have shown that every orbit has a representative $(m,n)$ satisfying (\ref{e:lat-cond}). The Proof of (ii) will show that there is only one such representative.

\fl{\bf Proof of (ii).} We will now argue that if $(m,n)$ satisfies (\ref{e:lat-cond}) then all other points on ${\cal O}_{(m,n)}$ either violate (\ref{e:lat-cond}) or coincide with $(m,n)$. Let us illustrate this with the point $(n-m,n) \in {\cal O}_{(m,n)}$. If this point lies in the wedge, then (\ref{e:lat-cond}) implies $n-m \ge 0$. But since $(m,n)$ already lies in the wedge we must also have $n \le \lfloor m/2 \rfloor$. Both conditions are satisfied only when $m=n=0$, in which case both points coincide: $(m,n) = (n-m,n) = (0,0)$. A similar argument applies to all other points in ${\cal O}_{m,n}$. This establishes that every orbit has a unique representative in the wedge defined by (\ref{e:lat-cond}). Thus, we may use $(m,n)$ satisfying (\ref{e:lat-cond}) to label a basis for the identity representation of $G$.

\small


\end{document}